\documentclass[a4paper,english,french]{article}
\usepackage{palatino}
\usepackage[T1]{fontenc}
\usepackage[latin1]{inputenc}
\usepackage{subfigure}
\usepackage{graphicx}
\usepackage{amssymb}

\makeatletter

\providecommand{\tabularnewline}{\\}

    \usepackage{renpar}
 \newcommand{\lyxaddress}[1]{
   \par {\raggedright #1 
   \vspace{1.4em}
   \noindent\par}
 }

\def\ggf{\hbox{\guillemotleft\ }}
\def\gdf{\hbox{\ \guillemotright}}

\usepackage{babel}
\makeatother
\begin{document}

\title{\vspace{-0.9cm}Prédiction de Performances pour les Communications
Collectives}

\author{Luiz Angelo Barchet-Estefanel%
\footnote{Financé par une bourse CAPES - Brésil BEX 1364/00-6%
}, Grégory Mounié}

\maketitle

\lyxaddress{Équipe MOAIS (CNRS-INPG-INRIA-UJF)\\
Laboratoire ID - IMAG\\
51, Avenue Jean Kuntzmann\\
38330 MONTONNOT SAINT MARTIN, France\\
\{Luiz-Angelo.Estefanel,Gregory.Mounie\}@imag.fr}

\Resume{Des travaux récents visent l'optimisation des opérations
de communication collective dans les environnements de type grille
de calcul. La solution la plus répandue est la séparation des communications
internes et externes à chaque grappe, mais cela n'exclut pas le découpage
des communications en plusieurs couches, pratique efficace démontrée
par Karonis \emph{et al.} \cite{key-53}. Dans les deux cas, la prédiction
des performances est un facteur essentiel, soit pour le réglage fin
des paramètres de communication, soit pour le calcul de la distribution
et de la hiérarchie des communications. Pour cela, il est très important
d'avoir des modèles précis des communications collectives, lesquels
seront utilisés pour prédire ces performances. Cet article décrit
notre expérience sur la modélisation des opérations de communication
collective. Nous présentons des modèles de communication pour différents
patrons de communication collective comme \ggf un vers plusieurs\gdf ,
\ggf un vers plusieurs personnalisé\gdf ~et \ggf plusieurs vers
plusieurs\gdf . Pour évaluer la précision des modèles, nous comparons
les prédictions obtenues avec les résultats des expérimentations effectuées
sur deux environnements réseaux différents, Fast Ethernet et Myrinet.}

\MotsCles{Communication Collective, Modèles de Communication, Prédiction
de Performance, MPI}

\section{\label{sec:Introduction}Introduction}

Plusieurs travaux récents visent l'implantation des opérations de
communication collective adaptées aux systèmes à grande échelle, notamment
les grilles. Dans ces environnements, l'hétérogénéité est un facteur
prépondérant qui doit obligatoirement être pris en compte \cite{key-55}.
Cette hétérogénéité représente, néanmoins, un vrai défi pour la prédiction
des performances, car les facteurs qui influencent les communications
ont des origines très variées, comme la distribution des processus
(par exemple, sur une grappe de machines multiprocesseurs), la distance
entre les machines et/ou les grappes, le taux d'utilisation du matériel
(surtout la congestion du réseau) et la variation de performance du
matériel. En effet, très souvent les grilles de calcul combinent différentes
machines et réseaux.

L'hétérogénéité inhérente à ces environnements, associée à la volatilité
des noeuds dans les grilles de calcul, empêche la création d'opérations
spécifiques pour ces environnements, comme en attestent \cite{key-2}
et \cite{key-52}. Pour simplifier cette modélisation, la plupart
des solutions considèrent les grilles comme l'interconnexion d'îlots
de grappes homogènes \cite{key-55}. Dans ce contexte, la majorité
des systèmes concentre l'optimisation au niveau des communications
entre les grappes, puisque ces liaisons sont généralement plus lentes
que celles intérieures à la grappe. Quelques exemples de cette approche
en deux couches incluent les bibliothèques ECO \cite{key-5}, MagPIe
\cite{key-6}\cite{key-9}, et même la bibliothèque LAM-MPI 7 \cite{key-32},
qui considère les machines SMP comme des îlots de communication rapide.
Il reste, néanmoins, la nécessité de régler \textbf{}les paramètres
de communication pour avoir des performances optimales, et pour cela,
la prédiction des performances à travers des modèles de communications
est un choix très avantageux.

Il existe, toutefois, la possibilité d'organiser les communications
en un plus grand nombre de couches. En effet, le travail de Karonis
et \emph{al.} \cite{key-53}\cite{key-31} a démontre que le découpage
en plusieurs couches de communication peut conduire à des réductions
du temps d'exécution plus importantes qu'un découpage en deux couches,
mais pour cela, il est nécessaire la connaissance \emph{a priori}
du coût de communication interne à chaque grappe. Dans ce cas, le
calcul de la distribution et de la hiérarchie des communications dépend
des temps de communication à l'intérieur des grappes, qui varient
selon l'opération de communication collective, le nombre de noeuds
et les caractéristiques du réseau de chaque grappe. 

D'autre part, la prédiction des performances des opérations collectives
est aussi intéressante pour d'autres environnements que les grilles.
En fait, même si on dispose d'une seule grappe, l'ordre d'exécution
des tâches peut influencer largement la performance des systèmes.
Dans ces cas, des travaux comme \cite{key-48} s'intéressent à la
prédiction du temps d'exécution d'une tâche, et pour cela, la connaissance
des performances de communication représente une étape très importante.

L'approche choisie pour ce travail est la prédiction des performances
à partir de la modélisation des opérations de communication collective,
par opposition aux prédictions fondées sur des expérimentations réelles
(voir Vadhiyar \emph{et al.} \cite{key-52}). Notre choix s'appuie
sur le fait que les prédictions obtenues à partir des modèles de communication
ont un coût très réduit par rapport aux expérimentations réelles,
sans pour autant perdre en précision. En effet, le travail de Vadhiyar
s'oriente maintenant vers la modélisation des performances pour réduire
le coût trop élevé des mesures pratiques \cite{key-10}.

Pour illustrer notre approche, ce travail présente des expériences
avec les opérations \emph{Broadcast}, \emph{Scatter} et \emph{All-to-All},
lesquelles représentent respectivement les patrons de communications
collectives \ggf un vers plusieurs\gdf ~(\emph{one-to-many}), \ggf un
vers plusieurs personnalisé\gdf ~(\emph{personalised one-to-many})
et \ggf plusieurs vers plusieurs\gdf ~(\emph{many-to-many}). Conceptuellement
simple, les patrons \ggf un vers plusieurs\gdf ~et \ggf un vers
plusieurs personnalisé\gdf ~sont aussi présents sur d'autres opérations
comme \emph{Barriers}, \emph{Reduces} et \emph{Gathers}. En revanche,
le patron \ggf plusieurs vers plusieurs\gdf est beaucoup plus complexe
parce qu'une opération comme \emph{All-to-All} est sujette à des importants
problèmes de congestion réseau.

Cet article présente notre expérience dans la construction de modèles
de performance qui caractérisent ces patrons de communication collective
très représentatifs. \foreignlanguage{english}{}Ces modèles sont utilisés
pour prédire la performance des opérations, mais aussi pour choisir
la technique d'implantation qui est la mieux adaptée à chaque ensemble
de paramètres (nombre de processus, taille des messages, performances
du réseau). Pour mieux démontrer l'efficacité de ces modèles, nous
avons exécuté des expérimentations sur deux environnements réseau
différents, Fast Ethernet et Myrinet.

La suite de cet article est organisée de la façon suivante : la Section
\ref{sec:System_Model} présente les définitions qui seront utilisées
dans cet article, ainsi que l'environnement de test. Les Sections
\ref{sec:Broadcast}, \ref{sec:Scatter} et \ref{sec:Alltoall} présentent
respectivement les modèles de communication développés pour les opérations
\emph{Broadcast}, \emph{Scatter} et \emph{All-to-All}, et aussi comparent
les prédictions des modèles avec les résultats obtenus à partir de
nos expériences. Finalement, la Section \ref{sec:Conclusions} présente
les conclusions et les perspectives futures de notre recherche.

\section{\label{sec:System_Model}Modèles et Définitions}

Pour créer des modèles précis de communications collectives, il est
souhaitable d'avoir un bon modèle de performance pour représenter
les communications bipoints. Dans le domaine des applications parallèles
avec échange de message, les modèles les plus utilisés sont \emph{BSP}
\cite{key-39} et \emph{LogP} \cite{key-3}. Même si ces deux modèles
sont équivalents dans la plupart des cas, \emph{LogP} est légèrement
plus général que BSP puisqu'il n'a pas besoin de barrières globales
qui séparent les phases de communication et calcul, mais aussi parce
que \emph{LogP} contient la notion de réseau de capacité finie, où
seulement un certain nombre de messages en transit sont supportés
simultanément \cite{key-42}. Comme conséquence, nous avons choisi
pour ce travail le modèle \emph{parameterised LogP (pLogP)} \cite{key-9}.
Le modèle \emph{pLogP} est une extension du modèle \emph{LogP} qui
peut traiter avec précision les petits comme les grands messages,
avec un minimum de complexité. À cause de cette simplicité, ce modèle
permet un prototypage rapide des opérations de communication collective,
et les modèles développés avec \emph{pLogP} ont permis la prédiction
des performances des communications avec une précision suffisante
dans la plupart des cas présentés.

\begin{figure}
\vspace{-0.3cm}\begin{tabular}{cc}
\subfigure[Fast Ethernet]{\includegraphics[%
  width=0.45\linewidth]{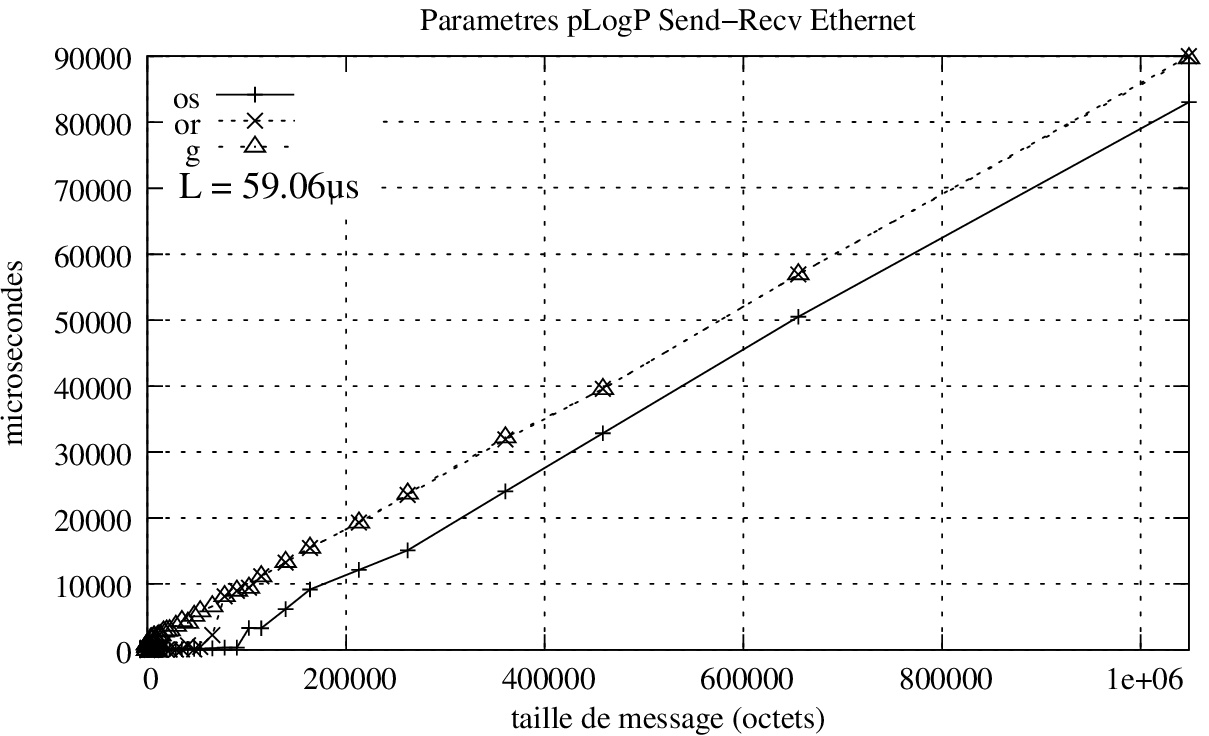}}&
\subfigure[Myrinet]{\includegraphics[%
  width=0.45\linewidth]{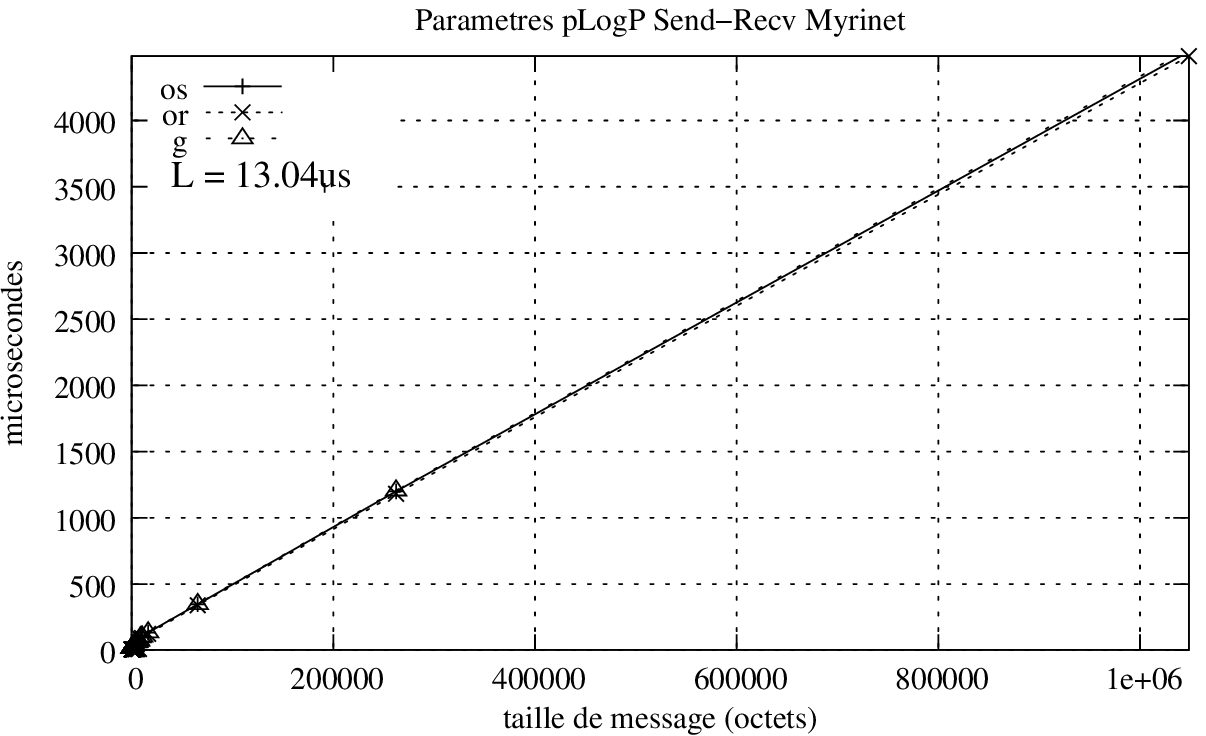}}\tabularnewline
\end{tabular}

\caption{\label{fig:pLogP-parameters}Paramètres \emph{pLogP} pour le réseau
icluster-2}
\end{figure}

Par conséquent, la terminologie employée dans cet article utilise
\textbf{\emph{g(m)}} pour représenter le coût d'envoi \textbf{}d'un
message de taille \emph{m} (le \emph{gap}), \textbf{\emph{os(m)}}
et \textbf{\emph{or(m)}} pour représenter respectivement le surcoût
dû à l'envoi et à la réception d'un message de taille \emph{m}, \textbf{\emph{L}}
pour représenter la latence entre deux noeuds, et \textbf{\emph{P}}
pour représenter le nombre de noeuds. Dans les cas où il y a segmentation
des messages, le segment de taille \emph{s} d'un message \emph{m}
est un multiple de la taille du type basique de données qui est transmis,
divisant alors le message initial \emph{m} en \emph{k} segments. Similairement,
\emph{g(s)} représente le \emph{gap} d'un segment de taille \emph{s}.
Ces paramètres furent obtenus avec l'outil MPI LogP Benchmark \cite{key-7},
et sont présentés sur la Figure \ref{fig:pLogP-parameters}.

Les expérimentation pratiques ont été conduites sur la grappe \textbf{icluster-2}
au centre de calcul de l'INRIA Rhône-Alpes%
\footnote{http://i-cluster2.inrialpes.fr/%
}. Cette grappe contient 104 ordinateurs Itanium-2 (IA-64, biprocesseur,
900MHz, 3GB) interconnectés pour des réseaux Fast Ethernet commuté
et Myrinet. Le système d'exploitation est Red Hat Linux Advanced Server
3.0 avec le noyau version 2.4.21smp. Les expérimentations utilisent
la bibliothèque LAM-MPI 7.0.4 \cite{key-32} et consistent en 100
mesures pour chaque ensemble de paramètres (taille du message, numéro
de processus), dont la valeur moyenne est considérée dans cet article.

Les prochaines sections détaillent les modèles de communication développés
pour les patrons de communication \ggf un vers plusieurs\gdf , \ggf un
vers plusieurs personnalisé\gdf ~et \ggf plusieurs vers plusieurs\gdf ,
ainsi que la validation de ces modèles à partir des expérimentations
pratiques.

\section{\label{sec:Broadcast}Un vers Plusieurs : \emph{Broadcast}}

Une opération de \emph{Broadcast} s'effectue quand un seul processus,
appelé \emph{racine,} envoie le même message de taille \emph{m} à
tous les autres $(P-1)$ processus. Des implantations classiques du
\emph{Broadcast} utilisent des arbres qui sont décrits par deux paramètres,
\emph{d} et \emph{h}, où \emph{d} est le nombre maximum de successeurs
qu'un noeud peut avoir, et \emph{h} est la hauteur de cet arbre, le
chemin le plus long qui relie la racine et les feuilles de cet arbre.
Un arbre optimal peut être construit à partir des paramètres du réseau
et avec \emph{d, h $\in$}{[}1...\emph{P}-1{]} tel que $\sum_{i=o}^{h}d^{i}\geq P$
est respecté, mais la plupart des implantations MPI utilisent deux
formes fixes, un Arbre Plat pour un nombre réduit de noeuds (jusqu'à
3 noeuds), et un Arbre Binomial pour un plus grand nombre de noeuds.

En plus de ces deux formes d'arbres, différentes techniques sont parfois
appliquées pour augmenter leur efficacité. Ces techniques peuvent
s'appliquer, pour exemple, à des grands messages, où un message de
\emph{rendez-vous} est envoyé pour préparer le récepteur afin de diminuer
les copies mémoires. On peut aussi utiliser des primitives de communication
non bloquantes pour permettre le recouvrement des communications et
du calcul. Malheureusement, ces techniques permettent juste des petites
améliorations, et la performance des communications reste néanmoins
liée aux caractéristiques du réseau.

Une autre possibilité de construire un \emph{Broadcast} est la composition
des chaînes de retransmission \cite{key-33}. Cette stratégie, utilisée
avec la segmentation des messages, présente des avantages importants,
comme l'indiquent \cite{key-9}\cite{key-34}\cite{key-54}. Dans
un \emph{Broadcast} à Chaîne Segmentée, la transmission des messages
en segments permet \textbf{}le \textbf{}recouvrement de la transmission
d'un segment \emph{k} et la réception du segment \emph{k}+1, minimisant
le \emph{gap}.

Le choix de la taille des segments reste, néanmoins, dépendant des
caractéristiques du réseau. En fait, le coût des segments trop petits
est plus dû à l'en-tête qu'à son contenu, et à l'inverse, des segments
trop grands ne sont pas capables d'exploiter tout le débit du réseau.
La recherche de la taille de segment \emph{s} qui minimise le temps
de communication peut se faire en utilisant les modèles présentés
dans le Tableau \ref{table:bcast_models}. D'abord, on cherche une
taille de segment \emph{s} qui minimise le temps de communication
parmi $s=m/2^{i}\;\mathrm{pour}\; i\in[0\ldots log_{2}m]$. Ensuite,
on peut affiner la recherche de la taille optimale avec des heuristiques
comme le \ggf \emph{local hill-climbing}\gdf ~proposée pour Kielmann
\emph{et al.} \cite{key-9}.

Nous avons établi plusieurs modèles pour représenter les stratégies
de communication et leurs techniques associées, qui sont présentées
sur le Tableau \ref{table:bcast_models}. La majorité de ces modèles
sont clairement inefficaces, donc nous avons choisi pour cet article
les stratégies d'Arbre Binomial et Chaîne Segmentée. Ces stratégies
seront analysées en Section \ref{sub:Broadcast_Practical}, où seront
comparées les prédictions des modèles avec les résultats issus des
expérimentations pratiques.

\begin{table}
\begin{center}\begin{tabular}{|c|c|}
\hline 
\textbf{\tiny Stratégie}&
\textbf{\tiny Modèle de Communication}\tabularnewline
\hline
\hline 
{\tiny Arbre Plat}&
{\tiny $(P-1)\times g(m)+L$}\tabularnewline
\hline 
{\tiny Arbre Plat Rendez-vous}&
{\tiny $(P-1)\times g(m)+2\times g(1)+3\ \times L$}\tabularnewline
\hline 
{\tiny Arbre Plat Segmenté}&
{\tiny $(P-1)\times(g(s)\times k)+L$}\tabularnewline
\hline 
{\tiny Chaîne}&
{\tiny $(P-1)\times(g(m)+L)$}\tabularnewline
\hline 
{\tiny Chaîne Rendez-vous}&
{\tiny $(P-1)\times(g(m)+2\times g(1)+3\ \times L)$}\tabularnewline
\hline 
{\tiny Chaîne Segmentée (Pipeline)}&
{\tiny $(P-1)\times(g(s)+L)+(g(s)\times(k-1))$}\tabularnewline
\hline 
{\tiny Arbre Binaire}&
{\tiny $\leq\lceil log_{2}P\rceil\times(2\times g(m)+L)$}\tabularnewline
\hline 
{\tiny Arbre Binomial}&
{\tiny $\lfloor log_{2}P\rfloor\times g(m)+\lceil log_{2}P\rceil\times L$}\tabularnewline
\hline 
{\tiny Arbre Binomial Rendez-vous}&
{\tiny $\lfloor log_{2}P\rfloor\times g(m)+$}\tabularnewline
&
{\tiny $\lceil log_{2}P\rceil\times(2\times g(1)+3\times L)$}\tabularnewline
\hline 
{\tiny Arbre Binomial Segmenté}&
{\tiny $\lfloor log_{2}P\rfloor\times g(s)\times k+\lceil log_{2}P\rceil\times L$}\tabularnewline
\hline
\end{tabular}\end{center}

\caption{\label{table:bcast_models}Modèles de communication pour le \emph{Broadcast}}
\end{table}

\subsection{\label{sub:Broadcast_Practical}Résultats Pratiques}

\begin{figure}
\vspace{-0.5cm}\begin{tabular}{cc}
\subfigure[Fast Ethernet]{\includegraphics[%
  width=0.45\linewidth]{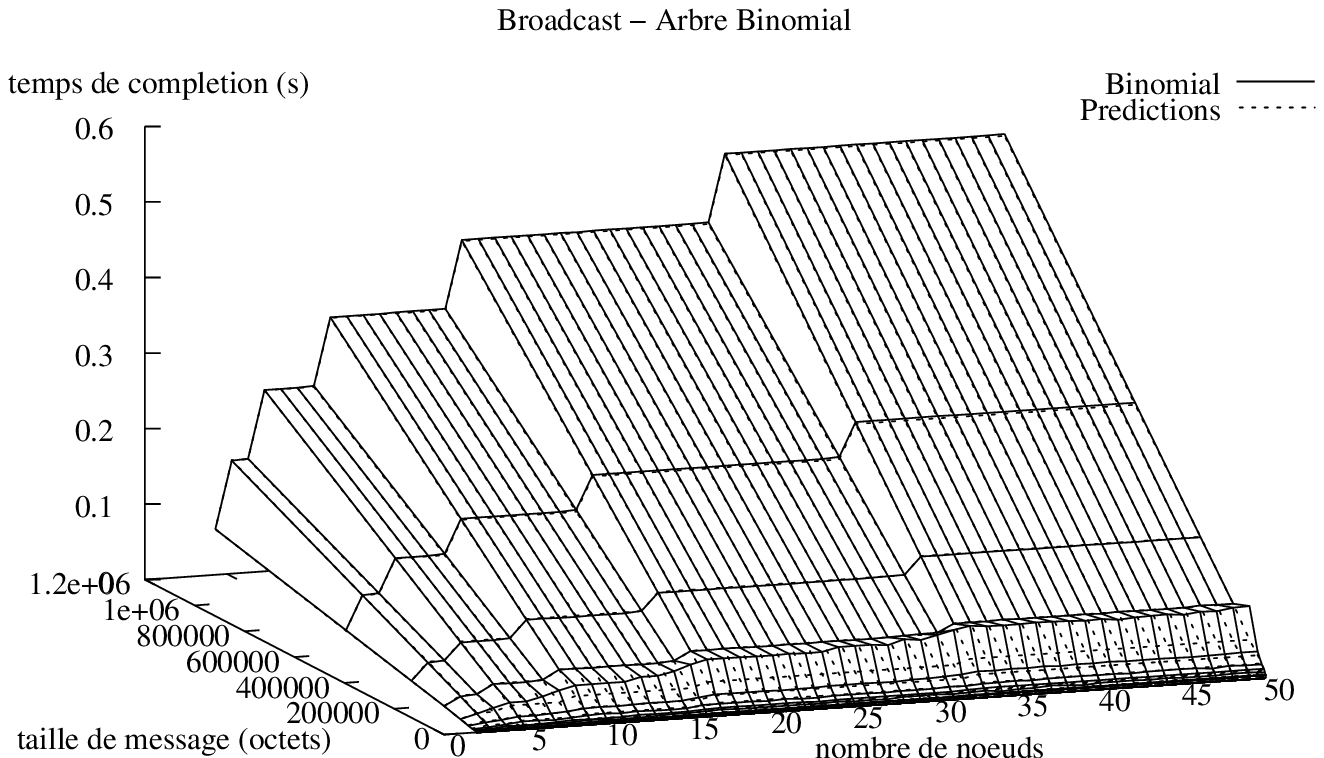}}&
\subfigure[Myrinet]{\includegraphics[%
  width=0.45\linewidth]{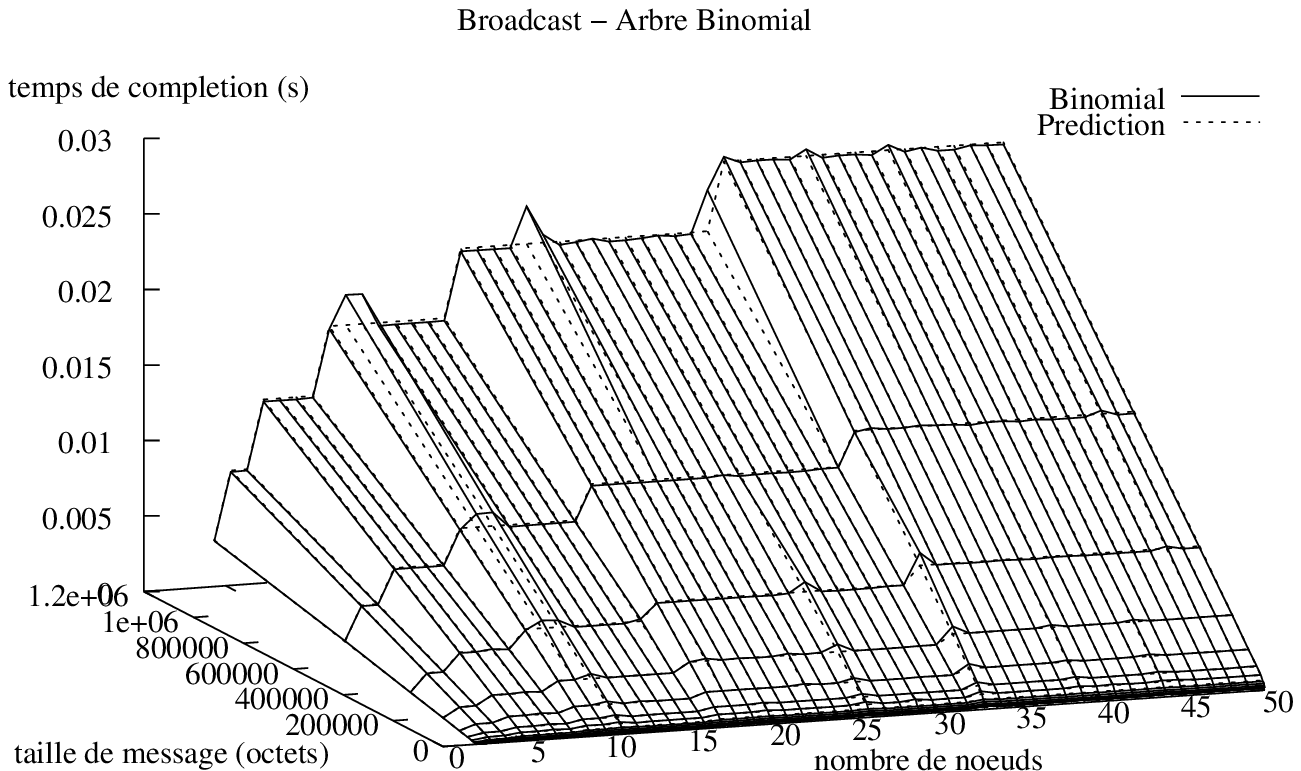}}\tabularnewline
\end{tabular}

\caption{\label{Figure:Comparison-Bcast_Bin}Les performances réelles et prédites
pour l'Arbre Binomial}
\end{figure}

\begin{figure}
\vspace{-0.5cm}\begin{tabular}{cc}
\subfigure[Fast Ethernet]{\includegraphics[%
  width=0.45\linewidth]{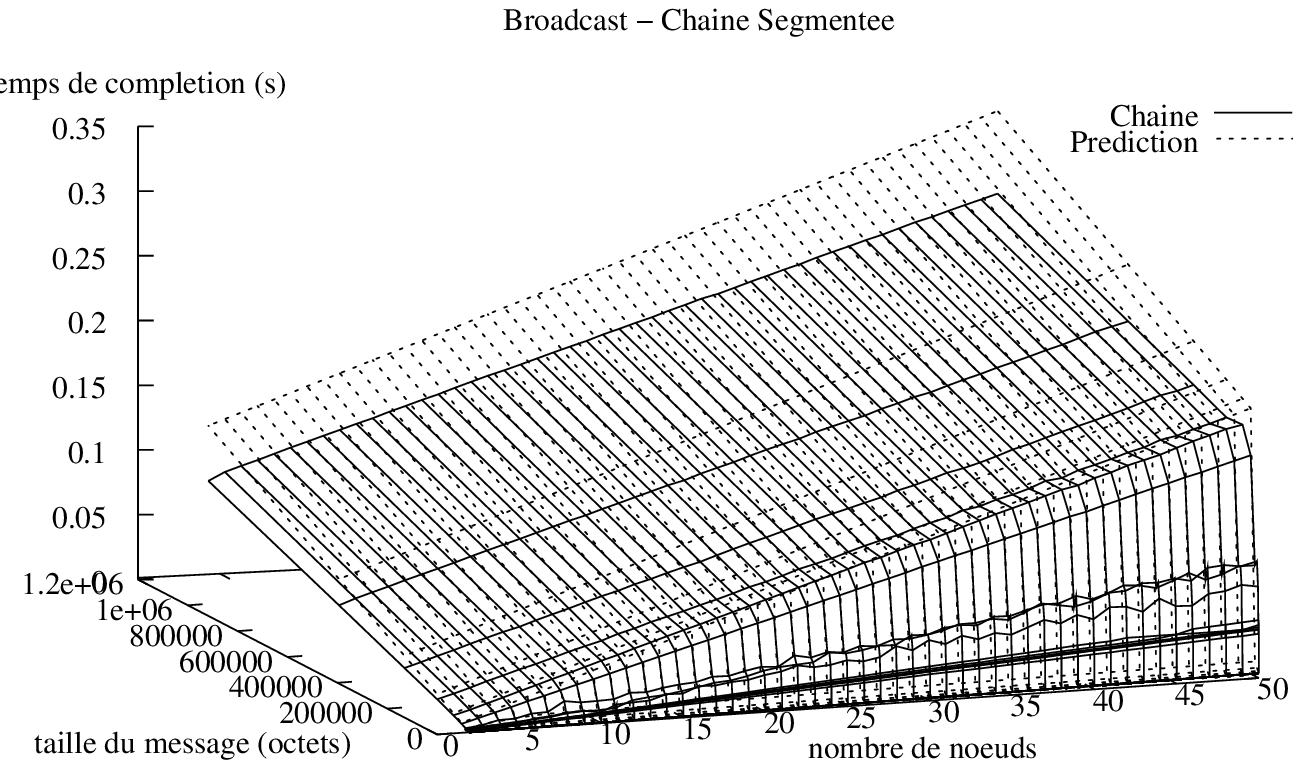}}&
\subfigure[Myrinet]{\includegraphics[%
  width=0.45\linewidth]{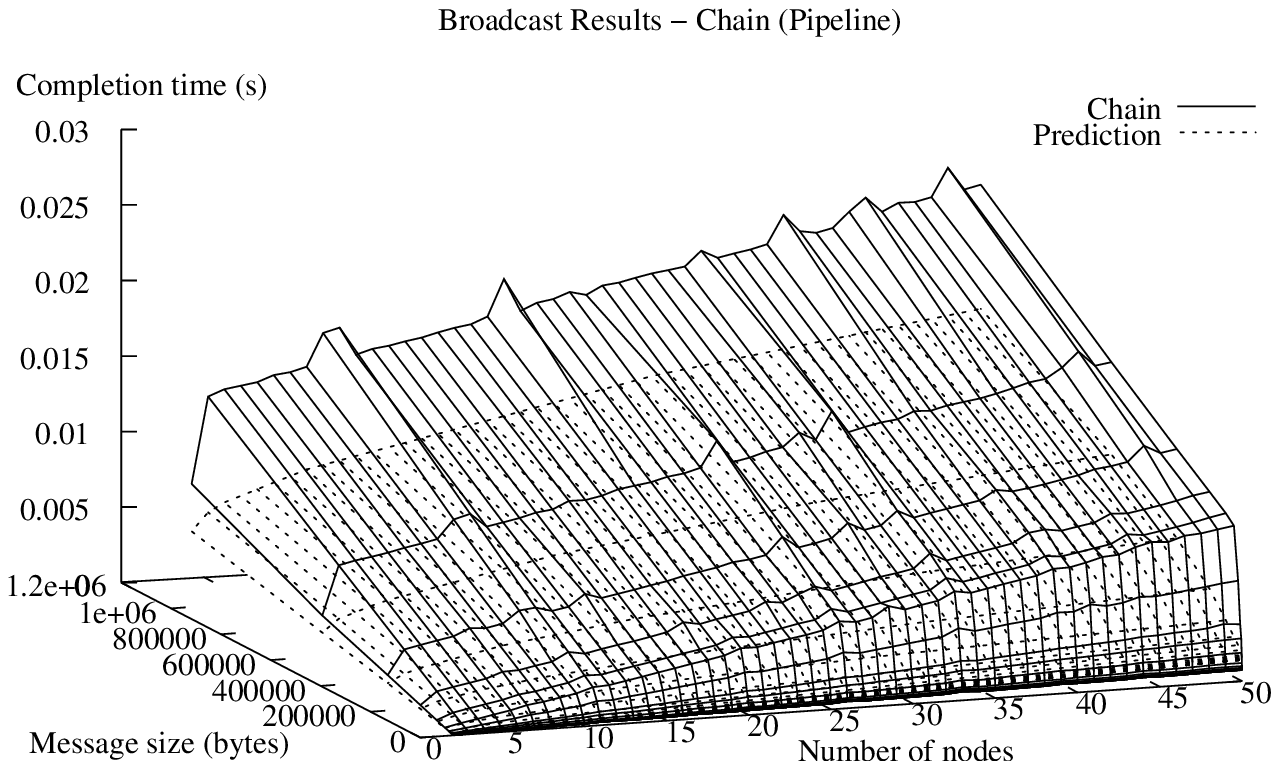}}\tabularnewline
\end{tabular}

\caption{\label{Figure:Comparison-Bcast-Chain}Les performances réelles et
prédites pour la Chaîne Segmentée}
\end{figure}

Pour évaluer la précision des modèles de communication, nous avons
obtenu les temps de communication des \emph{Broadcasts} en Arbre Binomial
et Chaîne Segmentée à partir des expériences pratiques, et ensuite
on les a comparés avec les prédictions des modèles. Les Figures \ref{Figure:Comparison-Bcast_Bin}
et \ref{Figure:Comparison-Bcast-Chain} présentent chaque stratégie
comparée avec les prédictions de son modèle. 

Les prédictions pour les Arbres Binomiaux (Figure \ref{Figure:Comparison-Bcast_Bin})
sont très proches des résultats pratiques. Pour la Chaîne Segmentée
(Figure \ref{Figure:Comparison-Bcast-Chain}), malgré les différences
entre les résultats réels et les prédictions, nous pouvons toujours
observer que les prédictions suivent le comportement des opérations
réelles. En effet, deux facteurs peuvent influencer fortement le résultat
de la Chaîne Segmentée: d'abord le coût de manipulation des segments
de message, et surtout la propagation des retards d'une machine à
toute la chaîne. Selon l'importance de ces deux facteurs, les résultats
obtenus seront plus ou moins éloignés du modèle de communication.

\begin{figure}
\vspace{-0.5cm}\begin{tabular}{cc}
\subfigure[Fast Ethernet]{\includegraphics[%
  width=0.45\linewidth]{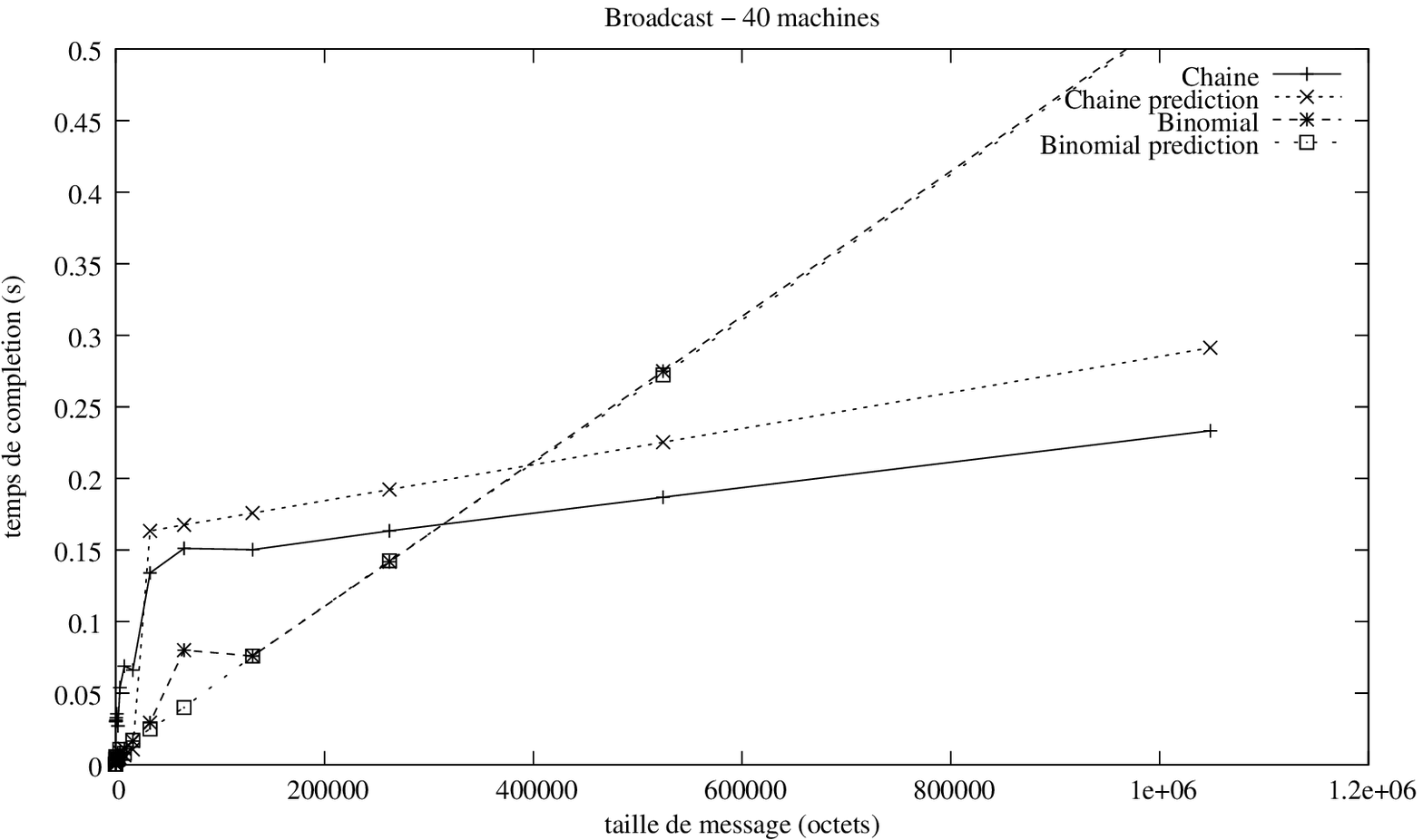}}&
\subfigure[Myrinet]{\includegraphics[%
  width=0.45\linewidth]{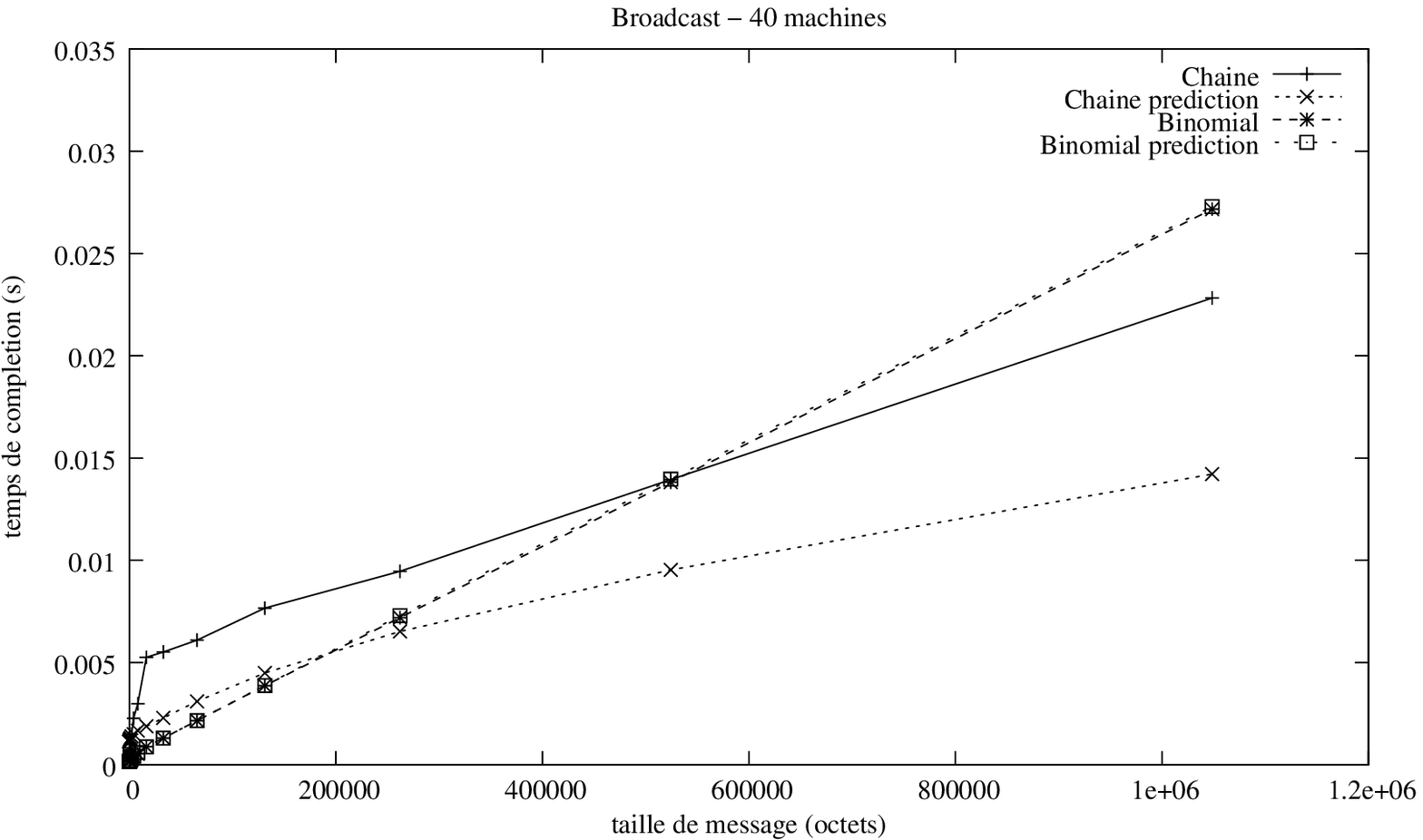}}\tabularnewline
\end{tabular}

\caption{\label{Figure:Comparison-between-models Bcast}Comparaison entre
les résultats réels et prédits pour un groupe de 40 machines}
\end{figure}

Finalement, la Figure \ref{Figure:Comparison-between-models Bcast}
compare directement les stratégies d'Arbre Binomial et Chaîne Segmentée
(et leurs prédictions) pour un groupe de 40 machines. Nous pouvons
observer que dans ce cas l'algorithme de Chaîne Segmentée est plus
performant pour des grands messages, même si les prédictions sont
moins précises que pour les autres modèles. 

Dans le cas des petits messages, des retards importants sont observés,
spécialement sur Ethernet. Les raisons de ces retards sont détaillées
dans plusieurs références, dont un article des développeurs du LAM-MPI
\cite{key-14}. Une enquête plus profonde, menée par Loncaric \cite{key-4},
a indiqué que ces retards sont dus à l'implantation des politiques
d'acquittement TCP sur Linux, qui occasionne le retard exceptionnel
d'un message à chaque \emph{n} messages transmis (dont \emph{n} dépend
de la version du noyau Linux). Ces retards sont observés même si l'option
socket TCP\_NODELAY est activée, et influencent fortement la performance
des Chaînes Segmentées à cause des tailles de segments utilisées.

\section{\label{sec:Scatter}Un vers Plusieurs Personnalisé : \emph{Scatter}}

L'opération \emph{Scatter}, aussi appelée \ggf \emph{broadcast} personnalisé\gdf ,
est une opération où le processus \emph{racine} détient \emph{$P$}
messages différents de taille \emph{$m$} qui seront distribués également
entre tous les P processus. Parce que le \emph{Scatter} est l'opération
symétrique de l'opération \emph{Gather}, les modèles développés pour
le \emph{Scatter} peuvent aussi représenter le patron de communication
\ggf plusieurs vers un\gdf ~de l'opération \emph{Gather}.

Dans le cas du \emph{Scatter}, où la \emph{racine} détient un message
différent pour chaque processus, il est généralement considéré que
le meilleur algorithme pour les réseaux homogènes utilise les Arbres
Plats \cite{key-9}. Par conséquent, l'implantation en Arbre Plat
est l'approche par défaut des bibliothèques MPI.

Ce choix est dû au fait que des alternatives pour les Arbres Plats
requièrent toujours que des grands ensembles de messages soient transmis
par des noeuds intermédiaires. En prenant par exemple le cas des Arbres
Binomiaux, le processus \emph{racine} transmet à ses successeurs des
paquets de messages qui contiennent plusieurs messages. Si d'un côté
cette stratégie peut bénéficier des envois parallèles, elle a des
inconvénients car la transmission des paquets de messages nécessite
plus de temps qu'un seul message. Par conséquent, l'efficacité des
Arbres Binomiaux dépend surtout de la vitesse de transmission des
grands messages, et on observe alors l'effet du compromis entre les
envois parallèles et la transmission des grands messages sur le temps
total de l'opération. 

Le Tableau \ref{table:Models-for-Scatter} présente les modèles de
communication développés pour le \emph{Scatter}. Pour ce travail,
on a choisi de comparer les approches des Arbres Plats et des Arbres
Binomiaux, plus performantes. Même si les Arbres Binomiaux ont un
surcoût dû à la transmission et manipulation des paquets de messages,
la possibilité de faire des envois simultanés doit être évaluée. D'ailleurs,
le modèle pour l'Arbre Binomial inclut la relation de compromis entre
le coût de transmission et les envois parallèles, ce qui nous donne
la possibilité d'évaluer les modèles en fonction des caractéristiques
du réseau.

\begin{table}
\begin{center}\begin{tabular}{|c|c|}
\hline 
\textbf{\scriptsize Stratégie}&
 \textbf{\scriptsize Modèle de Communication}\tabularnewline
\hline
{\scriptsize Arbre Plat}&
 {\scriptsize $(P-1)\times g(m)+L$}\tabularnewline
\hline
{\scriptsize Chaîne}&
 {\scriptsize $\sum_{j=1}^{P-1}g(j\times m)+(P-1)\times L$}\tabularnewline
\hline
{\scriptsize Arbre Binomial}&
 {\scriptsize $\sum_{j=0}^{\lceil log_{2}P\rceil-1}g(2^{j}\times m)+\lceil log_{2}P\rceil\times L$} \tabularnewline
\hline
\end{tabular}\end{center}

\caption{\label{table:Models-for-Scatter}Modèles de Communication pour le
\emph{Scatter}}
\end{table}

\subsection{\label{sub:Scatter_Practical}Résultats Pratiques}

\begin{figure}
\vspace{-0.5cm}\begin{tabular}{cc}
\subfigure[Fast Ethernet]{\includegraphics[%
  width=0.45\linewidth]{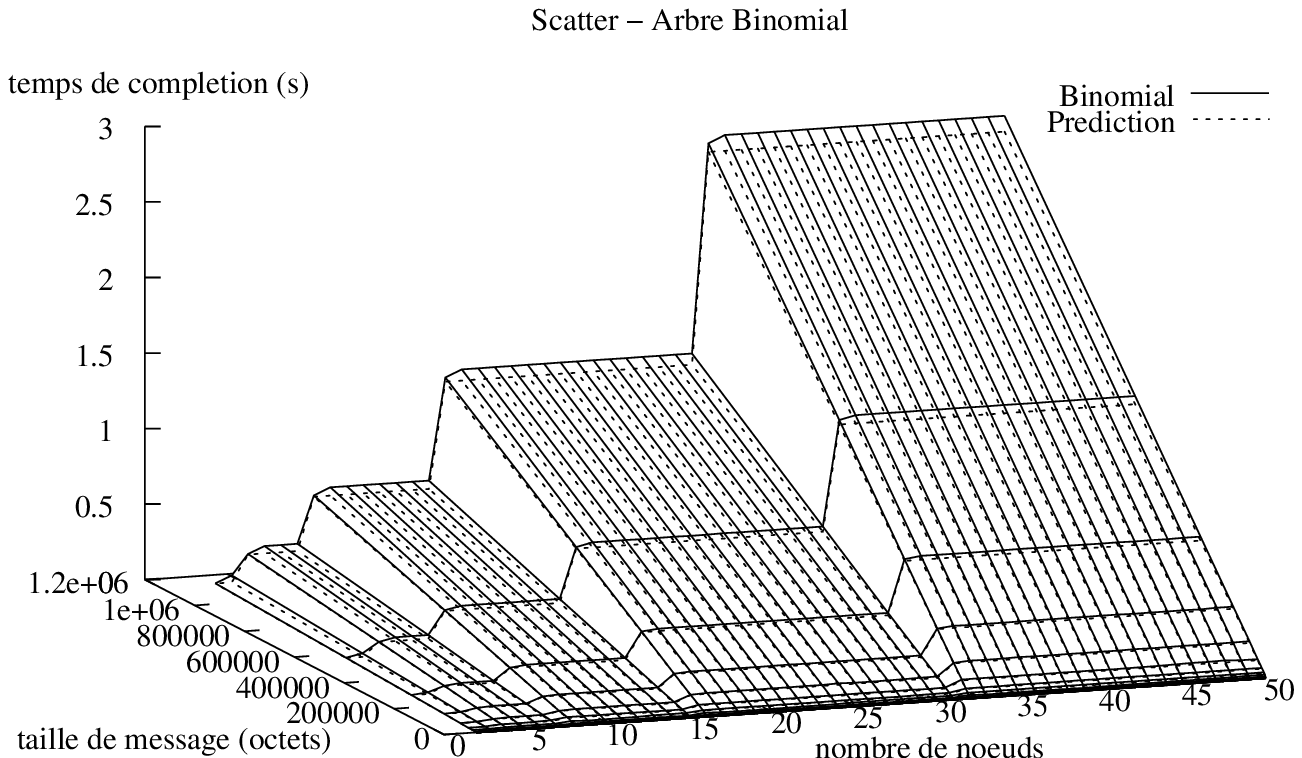}}&
\subfigure[Myrinet]{\includegraphics[%
  width=0.45\linewidth]{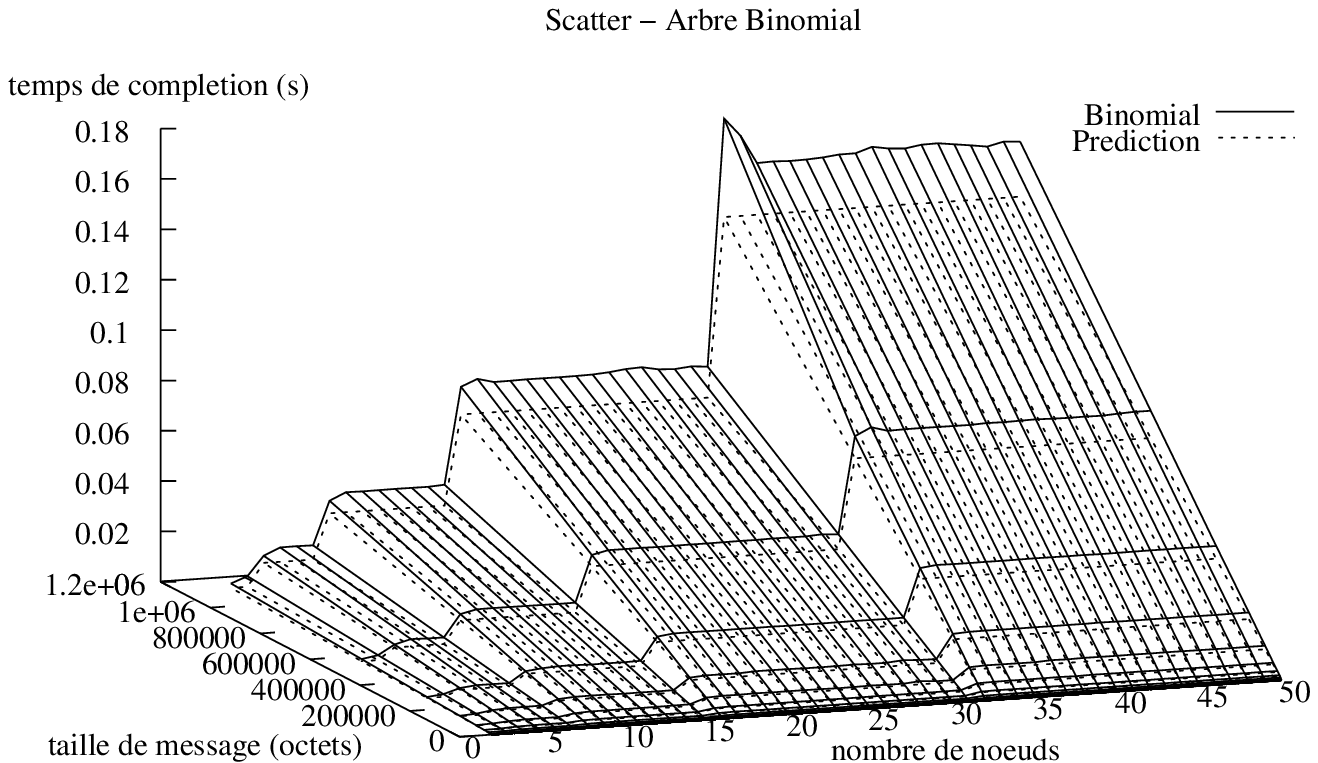}}\tabularnewline
\end{tabular}

\caption{\label{Figure:Comparison-Scatter-Bin}Performances réelles et prédites
pour le \emph{Scatter} en Arbre Binomial}
\end{figure}

\begin{figure}
\vspace{-0.5cm}\begin{tabular}{cc}
\subfigure[Fast Ethernet]{\includegraphics[%
  width=0.45\linewidth]{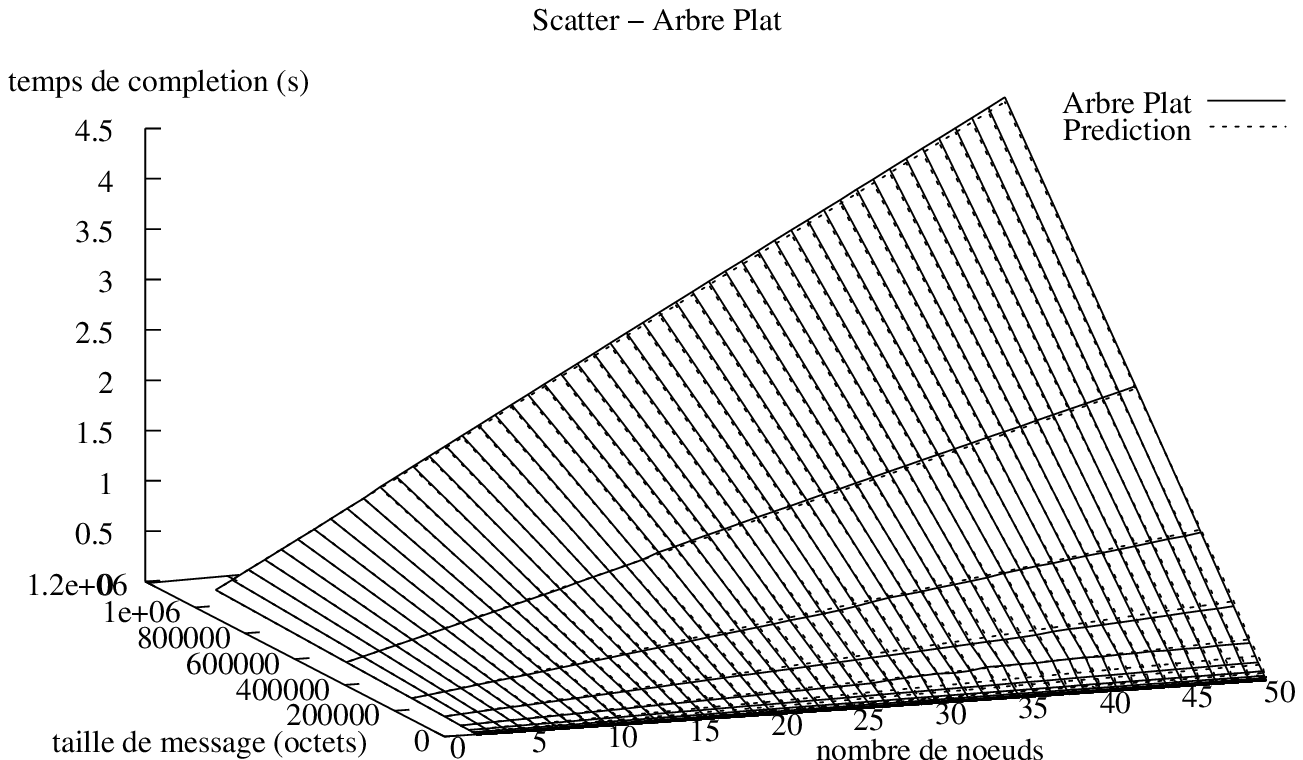}}&
\subfigure[Myrinet]{\includegraphics[%
  width=0.45\linewidth]{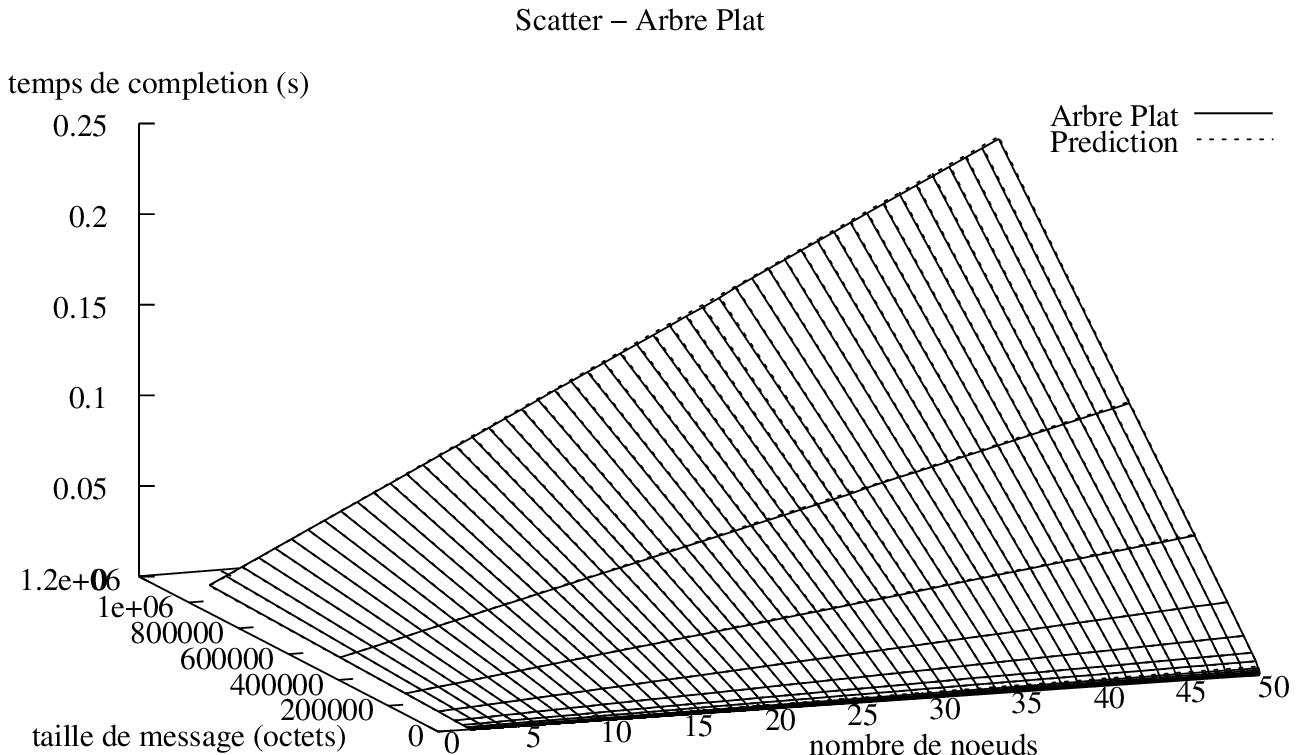}}\tabularnewline
\end{tabular}

\caption{\label{Figure:Comparison-Scatter-Flat}Performances réelles et prédites
pour le \emph{Scatter} en Arbre Plat}
\end{figure}

Une comparaison entre les résultats pratiques et les prédictions des
modèles est présentée dans les Figures \ref{Figure:Comparison-Scatter-Bin}
et \ref{Figure:Comparison-Scatter-Flat}. Nous pouvons observer que
les prédictions des modèles sont assez proches des résultats pratiques.
Les différences observées dans le cas des Arbres Binomiaux sont plutôt
dues au coût de manipulation des paquets de messages (extraction,
sélection, repaquetage), qui n'est pas représenté par le modèle de
performance (\emph{pLogP}). 

À cause des caractéristiques de notre réseau, on observe que l'approche
des Arbres Binomiaux est fréquemment plus efficace que l'approche
des Arbres Plats. Plus exactement, la simplicité du modèle Arbre Plat
est supplantée par la capacité de répartir la charge des transmissions
entre plusieurs noeuds. Ce résultat s'avère très utile pour l'augmentation
des performances de l'opération \emph{Scatter}. 

\begin{figure}
\vspace{-0.5cm}\begin{tabular}{cc}
\subfigure[Fast Ethernet]{\includegraphics[%
  width=0.45\linewidth]{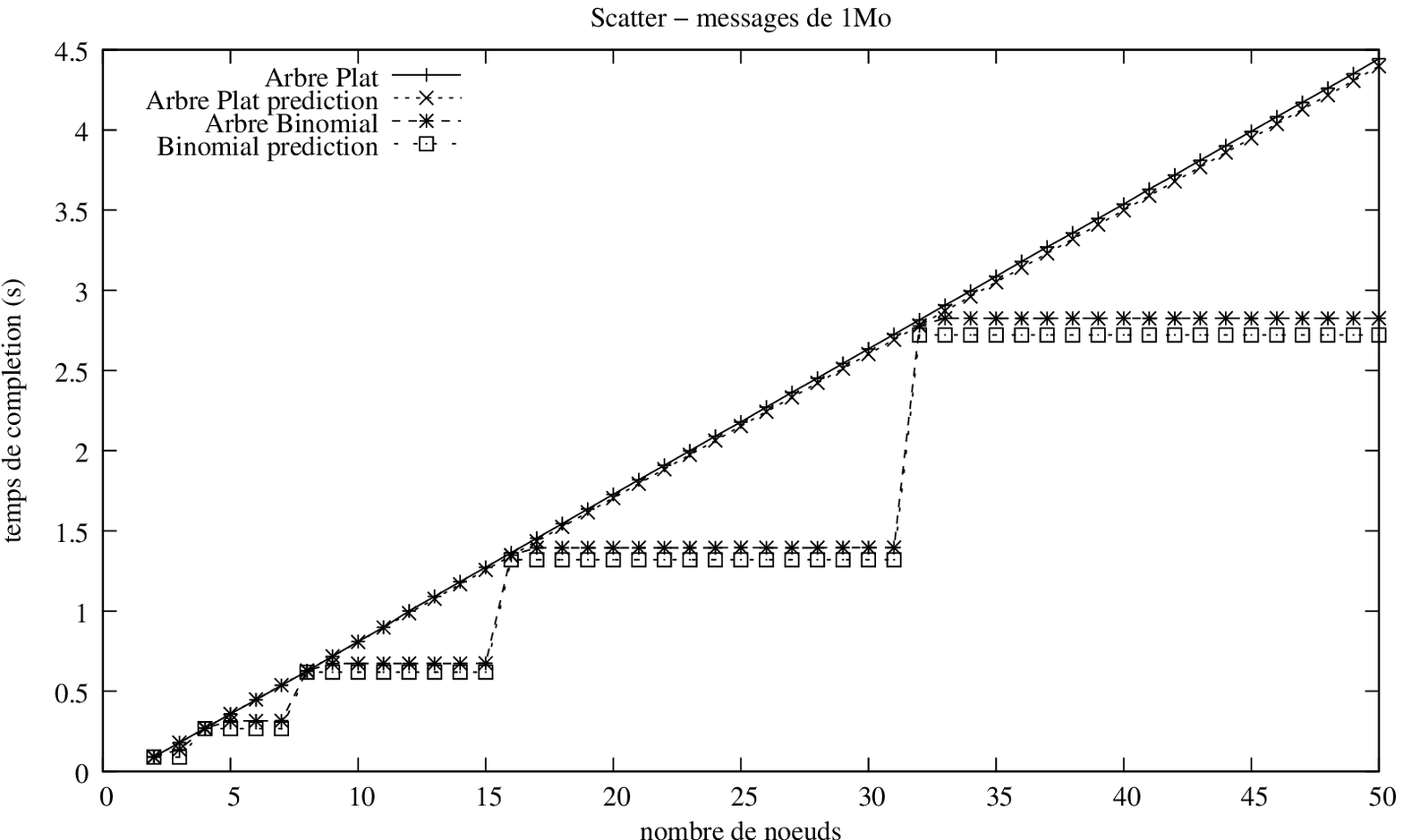}}&
\subfigure[Myrinet]{\includegraphics[%
  width=0.45\linewidth]{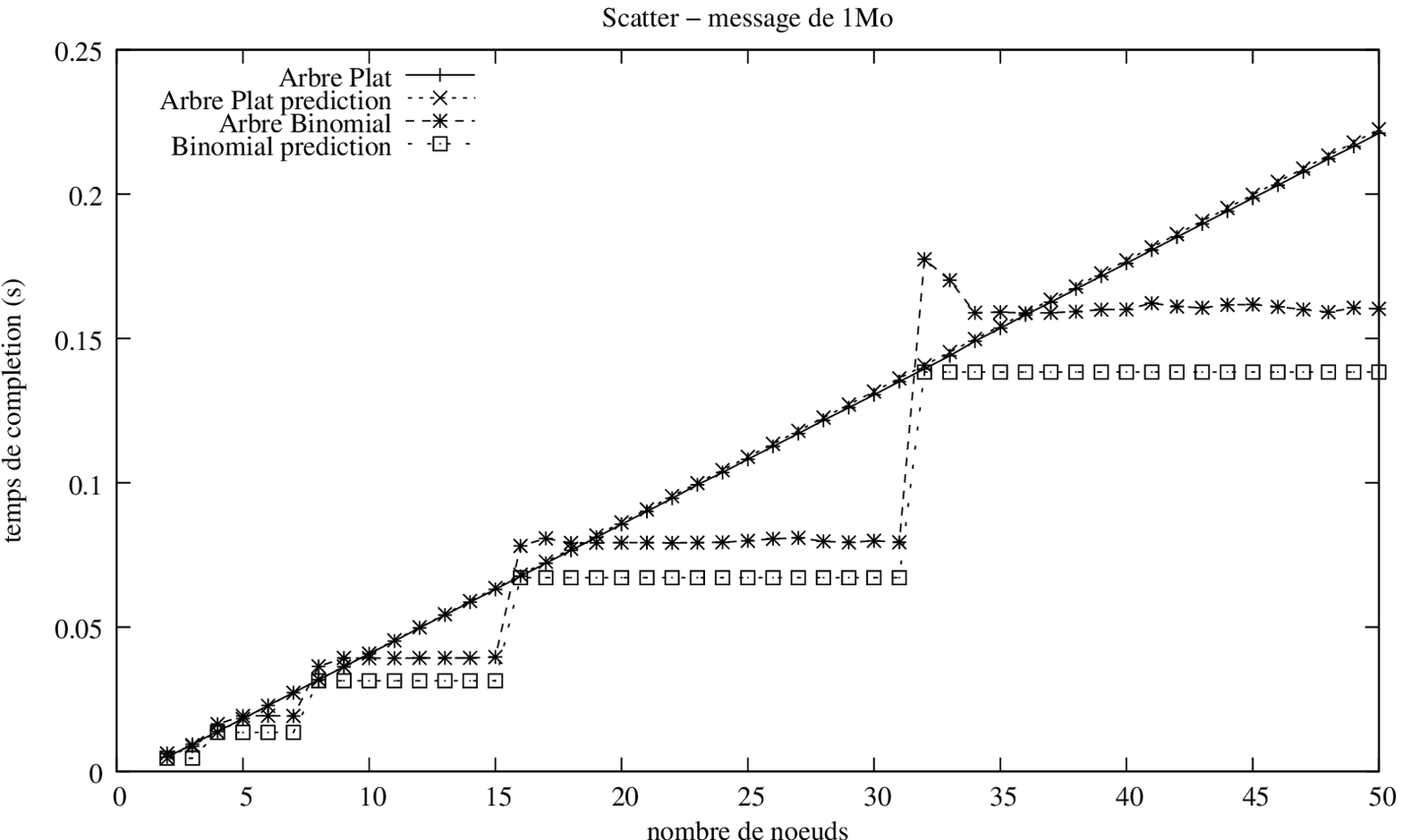}}\tabularnewline
\end{tabular}

\caption{\label{Figure:Comparison-between-models-Scatter}Comparaison entre
les résultats réels et prédits pour des messages de 1Mo}
\end{figure}

Lorsque le modèle en Arbre Plat est limité par le temps nécessaire
à la transmission des messages successifs (le \emph{gap}), sa performance
est directement liée au nombre de processus. En revanche, le coût
du modèle en Arbre Binomial augmente de façon logarithmique ({\small $\lceil log_{2}P\rceil$}),
ce qui offre des performances très avantageuses aux communications
avec un nombre de noeuds légèrement inférieur à $2^{x}\:\mathrm{pour}\: x\in\mathbb{N}$.
Néanmoins, la variation des performances du modèle en Arbre Binomial
selon le nombre de processus encourage la comparaison préalable des
modèles de performance, de manière à choisir l'algorithme qui s'adapte
le mieux à chaque ensemble de paramètres (taille de message, nombre
de noeuds), comme illustre la Figure \ref{Figure:Comparison-between-models-Scatter}.

\section{\label{sec:Alltoall}Plusieurs vers Plusieurs : \emph{All to All}}

Un des plus importants patrons de communication collective pour des
applications scientifiques est l'échange total \cite{key-35}, où
les algorithmes parallèles alternent des périodes de calcul avec des
périodes d'échange de données entre les processus. Pour cela, une
des opérations plus répandues est le \emph{All-to-All}, qui permet
la transposition des données appartenant à un groupe de processus.
Dans le cas de l'opération \emph{All-to-All}, chaque processus détient
\emph{$m\times P$} unités de données qui seront distribuées également
entre les P processus.

Plusieurs travaux visent l'optimisation du \emph{All-to-All} et sa
variante \emph{All-to-All-v}, qui permet l'envoi des messages avec
des tailles différentes pour chaque processus. Cependant, la plupart
des propositions sont adaptées à des structures d'interconnexion très
spécifiques, comme dans le cas des topologies en grille, tores et
hypercubes \cite{key-35}. Des solutions générales, comme celles implémentées
sur plusieurs distributions MPI, considèrent que chaque processus
ouvre une communication directe avec les autres processus.

\begin{figure}
\vspace{-0.3cm}\begin{tabular}{cc}
\subfigure[Fast Ethernet]{\includegraphics[%
  width=0.45\linewidth]{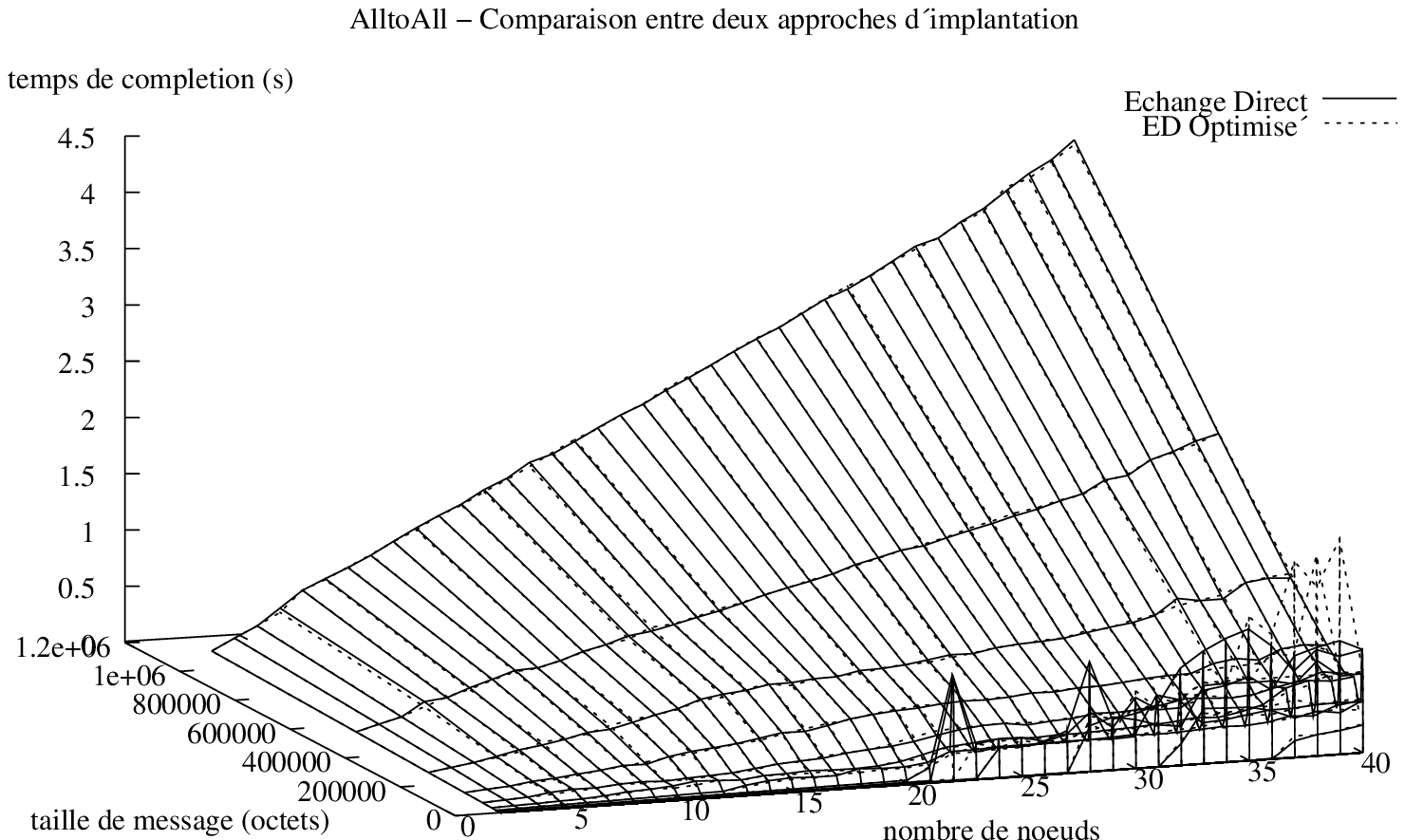}}&
\subfigure[Myrinet]{\includegraphics[%
  width=0.45\linewidth]{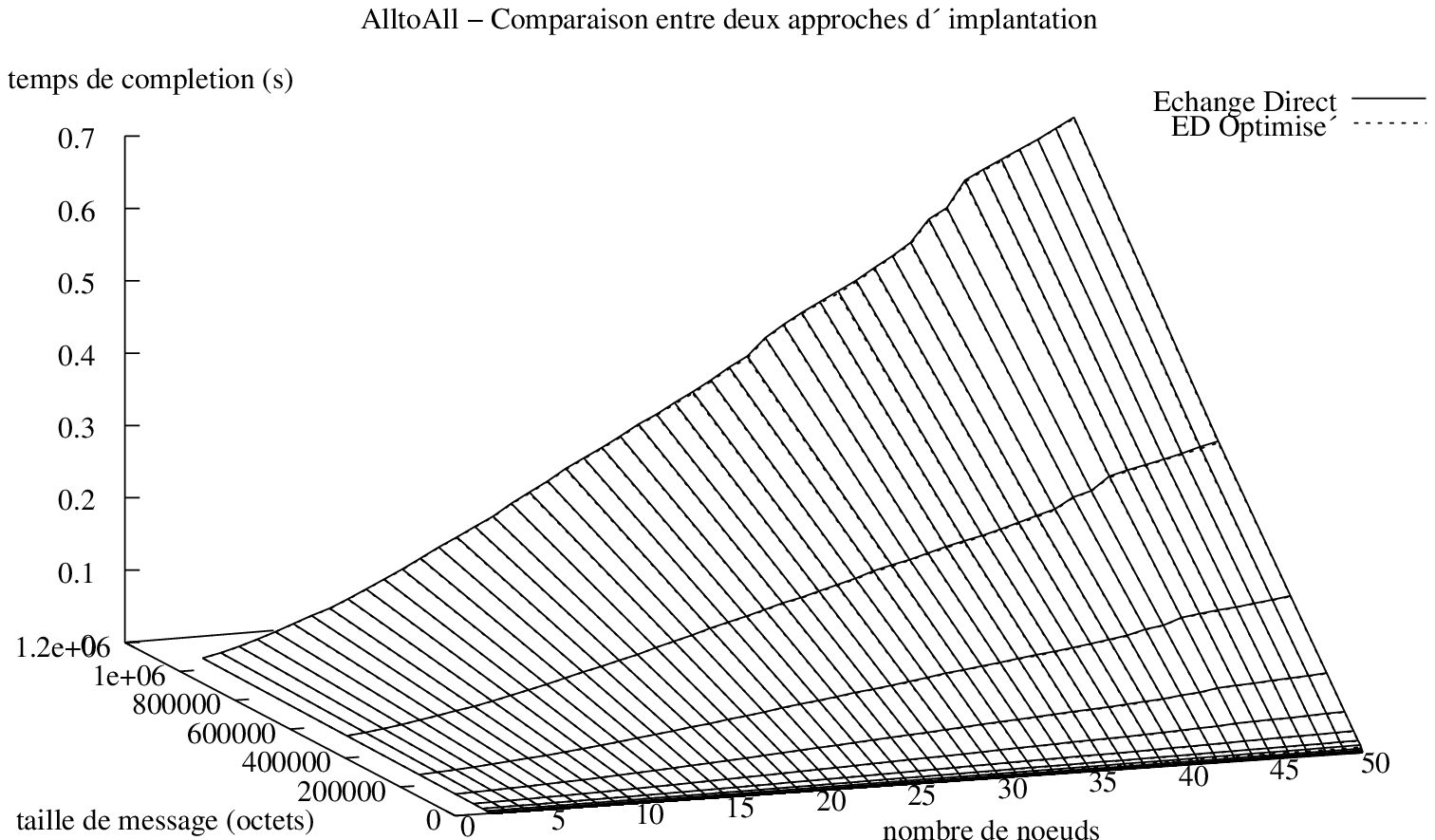}}\tabularnewline
\end{tabular}

\caption{\label{cap:Comparaison-DE}Comparaison entre les approches \emph{Échange
Direct} et \emph{Échange Direct Optimisé}}
\end{figure}

L'approche la plus simple d'implantation de \emph{All-to-All}, que
l'on appellera \emph{Échange Direct}, considère que chaque processus
communique directement avec les autres, et que tous les appels d'envois
et de réceptions sont initiés simultanément. Un exemple de l'approche
\emph{Échange Direct} est l'implantation de \emph{MPI\_Alltoall} de
LAM version 6.5.2 \cite{key-41}. Dû à ses caractéristiques, cet algorithme
peut avoir des problèmes de surcharge du récepteur, car les processus
suivent le même ordre d'envoi, surchargeant un seul processus récepteur
à chaque tour. À cause de cela, une optimisation simple consiste en
faire la rotation des listes de destinataires, comme le font déjà
les implantations MPI LAM 7.0.4 \cite{key-32} et MPICH 1.2.5 \cite{key-40}.
Malgré cette optimisation, des tests pratiques n'ont pas démontré
une grande influence sur le résultat, comme démontre la Figure \ref{cap:Comparaison-DE}.
Nos expériences suggèrent que la surcharge d'un récepteur est un problème
mineur en comparaison avec l'occurrence de la congestion réseau. En
fait, l'analyse faite par Grove \cite{key-36} indiquait déjà que
les ralentissements observés sont plutôt dus à des pertes de paquets
et leurs timeouts de retransmission TCP/IP causés par la surcharge
du réseau.

D'ailleurs, on observe sur la Figure \ref{cap:Comparaison-DE}(a)
des grandes variations de performance pour les messages petits. Ces
variations, observées seulement sur le réseau Fast Ethernet, sont
probablement dues aux problèmes de retard de petits messages déjà
discutés en Section \ref{sub:Broadcast_Practical}. Le fait que ces
retards sont plus importants que ceux observés dans les cas de l'opération
de \emph{Broadcast} reflète simplement le surcoût du patron \ggf plusieurs
vers plusieurs\gdf .

Par conséquent, la difficulté des modèles de communication pour le
\emph{All-to-All} réside dans la prise en compte des spécificités
du patron de communication \ggf plusieurs vers plusieurs\gdf . Des
modèles théoriques comme ceux présentés par \cite{key-35} sont pour
la plupart des simples extensions du modèle \emph{Scatter}, et ne
tiennent pas compte de l'influence de la congestion réseau, par exemple.

En fait, la plupart des travaux de modélisation de performance utilisent
des résultats des communications bipoints pour abstraire la performance
des communications collectives. Tam et Wang \cite{key-44}\cite{key-49}
ont démontré, toutefois, que le temps d'exécution des opérations de
communication collective, spécialement le \emph{Gather} et le \emph{All-to-All},
est fortement dominé par la congestion du réseau et par la perte de
paquets de messages, ce qui rend très difficile la quantification
de ces effets.

Pour sa thèse de doctorat, Grove \cite{key-36} a élaboré une étude
très intéressante à propos de modèles de performance. D'une manière
concise, son étude a présenté une vision riche du développement et
de l'évolution des modèles de performance. Ainsi, il démontre que
seulement ces dernières années la congestion est devenue une préoccupation
réelle, et que l'influence de la congestion reste encore un des grands
défis des concepteurs de modèles.

Un des premiers modèles qui considère les effets de la congestion
des ressources fut présenté par Adve \cite{key-51}. Ce modèle considérait
que le temps total d'exécution était réparti entre quatre composants
:

\vspace{-0.3cm}\begin{eqnarray*}
T & = & t_{computation}+t_{communication}+t_{resource-contention}+t_{synchronisation}\end{eqnarray*}

Malgré sa simplicité conceptuelle, ce modèle n'est pas trivial à cause
de la nature non-déterministe de la congestion, et surtout de la difficulté
à déterminer les retards moyens de synchronisation.

Même si la congestion des ressources est difficile à modéliser, Clement
et Steed \cite{key-46} ont introduit un moyen simple pour exprimer
la congestion sur des réseaux partagés, comme par exemple l'Ethernet
non commuté, qui consiste d'un facteur de congestion $\gamma$ qui
augmente un modèle de communication linéaire~T~:

\vspace{-0.1cm}\[
T=l+\frac{b\gamma}{W}\]

où \emph{l} est la latence du lien, \emph{b} est la taille du message,
W est le débit du lien et $\gamma$ représente le nombre de processus.
Ce modèle augmente la précision des prédictions avec un coût minimum,
mais pour cela il faut encore que tous les processus communiquent
simultanément, ce qui n'est pas vrai que pour quelques patrons de
communication.

Ce résultat est fortement lié au travail de Labarta, Girona et \emph{al.}
\cite{key-50}, qui tente d'approcher le comportement de la congestion
réseau au considérer que s'il y a \emph{m} messages à transmettre,
et seulement \emph{b} canaux disponibles, les messages sont sérialisés
en $\left\lceil \frac{m}{b}\right\rceil $ vagues de communication.

Certains modèles de performance orientés à la congestion sont apparus
récemment. \emph{LoGPC} \cite{key-47} est une extension du modèle
\emph{LogP} qui détermine l'influence de la congestion à travers l'analyse
des filles d'attente sur un réseau de \emph{n} cubes de dimension
\emph{k} chacun. Cette analyse rend très difficile l'utilisation pratique
du modèle. Une autre approche, plus pratique, est celle de Tam \cite{key-49},
qui considère la congestion comme part intégrante de la latence. Par
conséquent, ce modèle utilise des valeurs de latence qui varient selon
la taille du message. Si cette approche est beaucoup plus simple à
implémenter, le surcoût dû à l'obtention des valeurs de latence pour
plusieurs tailles de messages devient trop élevé quand on considère
des réseaux de longue distance. 

Pour ce travail nous adoptons une approche similaire à Clement et
Steed \cite{key-46}, où la congestion est suffisamment linéaire pour
être modélisée. Notre approche consiste à identifier le comportement
de l'opération \emph{All-to-All} par rapport à des performances théoriques
établies à partir du modèle de communication \emph{1-port}. Notre
hypothèse est que la congestion dépend plutôt des caractéristiques
physiques du réseau (cartes, liens, commutateurs, ...), de façon que
le rapport entre le résultat pratique et les performances théoriques
devient une \ggf signature\gdf ~de ce réseau. Une fois identifié
ce rapport, nous pouvons l'utiliser pour prédire la performance d'autres
exécutions effectuées sur le même réseau.

Dans le cas des communications du type \emph{All-to-All}, les valeurs
de performance théorique sont obtenues à partir de l'extension du
modèle \emph{Scatter}, mais cette fois-ci tenant compte des caractéristiques
et restrictions du patron \ggf plusieurs vers plusieurs\gdf , en
particulier la capacité des noeuds à recouvrir l'envoi et la réception
des messages. 

En fait, selon le modèle de communication \emph{1-port}, un processus
peut envoyer et recevoir des messages simultanément. Cependant, des
restrictions dues à la congestion peuvent forcer les machines à sérialiser
leurs envois et réceptions. Dans ce cas, en reprenant les notions
de \emph{pLogP}, nous étudions le fait que même si deux messages ne
peuvent pas être envoyés consécutivement en moins de \emph{g} unités
de temps à travers le même lien, il suffit de \emph{os} unités de
temps pour envoyer un message (plus spécifiquement, pour délivrer
le message à la carte réseau) et \emph{or} pour le recevoir. 

Par conséquent, la limite inférieure est représentée par la capacité
d'envoyer et recevoir des messages simultanément. Pour la limite supérieure
théorique, les noeuds sérialisent leurs envois et réceptions. Il est
possible qui la performance réelle dépasse la limite supérieure, car
existent d'autres facteurs qui peuvent influencer les communications.
Toutefois, l'observation des limites théoriques permet la séparation
des facteurs liés au transit des messages et les facteurs dus au matériel
physique, ce qui rend possible la définition de cette \ggf signature\gdf ~du
réseau. Les formules pour les limites théoriques sont présentées sur
le Tableau \ref{table:Models-for-Alltoall}.

\begin{table}
\begin{center}\begin{tabular}{|c|c|}
\hline 
&
\textbf{\scriptsize Modèle de Communication}\tabularnewline
\hline
{\scriptsize Limite Supérieure}&
 {\scriptsize $(P-1)\times os(m)+(P-1)\times or(m)+L$}\tabularnewline
\hline
{\scriptsize Limite Inférieure}&
 {\scriptsize $(P-1)\times g(m)+L$} \tabularnewline
\hline
\end{tabular}\end{center}

\caption{\label{table:Models-for-Alltoall}Limites de communication pour l'opération
\emph{All-to-All}}
\end{table}

\subsection{\label{sub:Alltoall-Practical-Results}Résultats Pratiques}

\begin{figure}
\vspace{-0.6cm}\begin{tabular}{cc}
\subfigure[Fast Ethernet]{\includegraphics[%
  bb=0bp 0bp 467bp 275bp,
  width=0.45\linewidth]{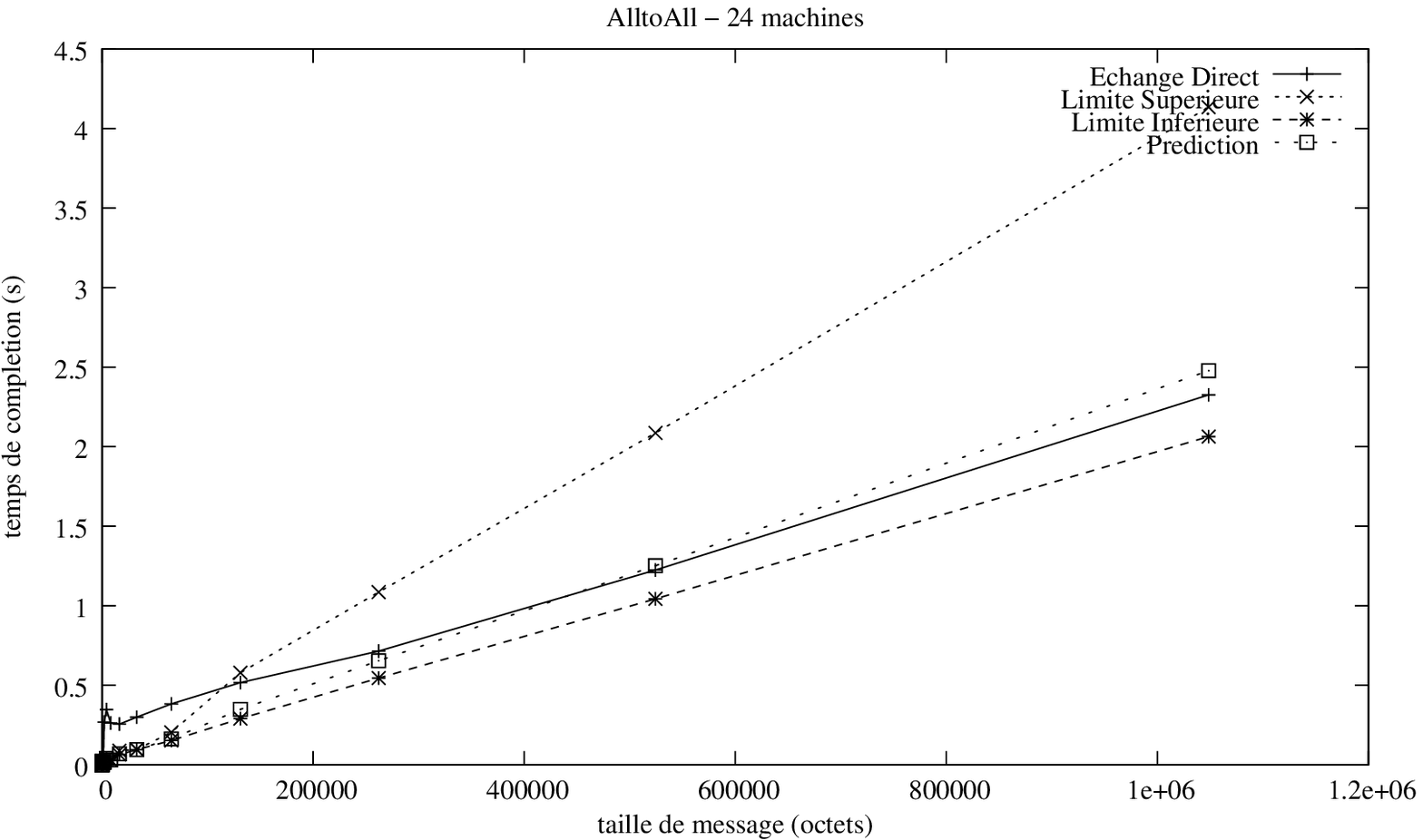}}&
\subfigure[Myrinet]{\includegraphics[%
  width=0.45\linewidth]{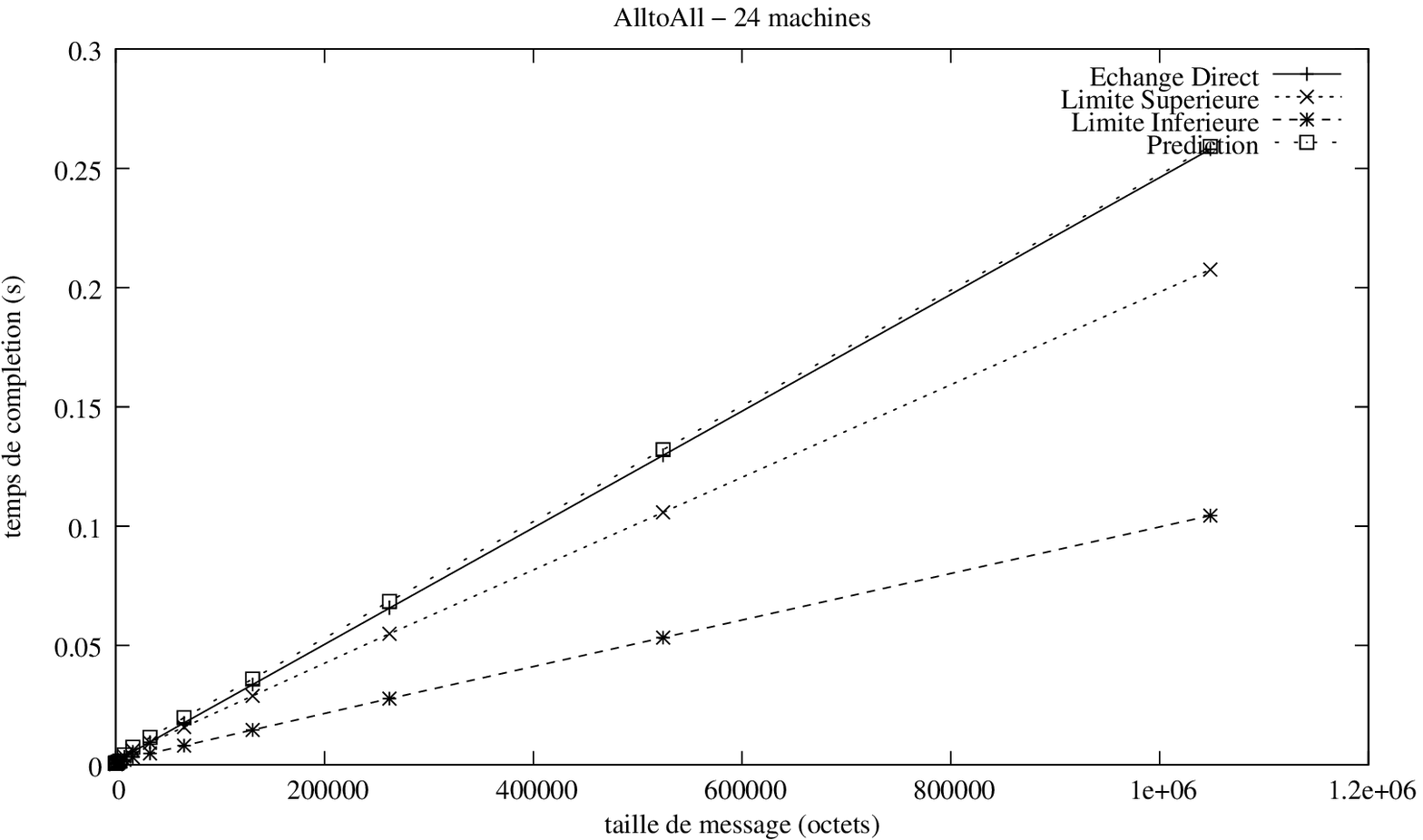}}\tabularnewline
\end{tabular}

\caption{\label{Figure:Comparison-bounds}Performance de l'algorithme \emph{All-to-All}
comparé aux limites théoriques et aux prédictions du modèle, pour
24 machines}
\end{figure}

Pour illustrer notre approche, nous présentons en Figure \ref{Figure:Comparison-bounds}
les résultats des expériences avec l'algorithme \emph{Échange Direct}
et les limites théoriques pour 24 machines. La première observation
importante est l'écart entre le résultat réel et la limite inférieure
(basé sur le modèle \emph{Scatter}). Cette différence non négligeable
est déjà due aux effets de la congestion du réseau. 

L'observation de ces valeurs permet l'approximation des résultats
réels à travers une relation de congestion établie entre les limites
théoriques. Cette relation de congestion $\gamma$ est constante et
dépend uniquement des caractéristiques du réseau, dont les limites
inférieures et supérieures (définies dans le Tableau \ref{table:Models-for-Alltoall})
dépendent du nombre de processus. Ainsi, nous proposons pour cette
relation la formule suivante :

\vspace{-0.2cm}\[
T=LimInferieure+(LimSuperieure-LimInferieure)\times\gamma\]

\begin{figure}
\vspace{-0.3cm}\begin{tabular}{cc}
\subfigure[Fast Ethernet]{\includegraphics[%
  bb=0bp 0bp 374bp 217bp,
  width=0.45\linewidth]{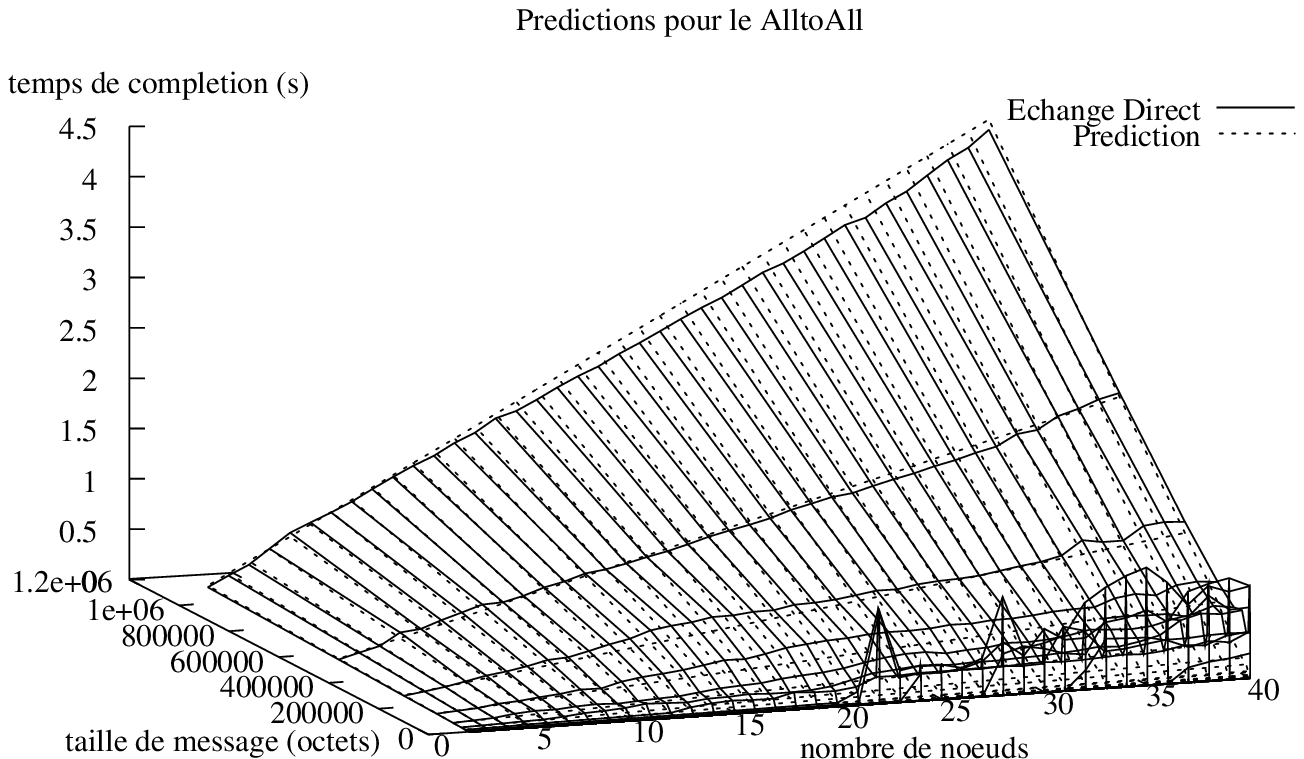}}&
\subfigure[Myrinet]{\includegraphics[%
  width=0.45\linewidth]{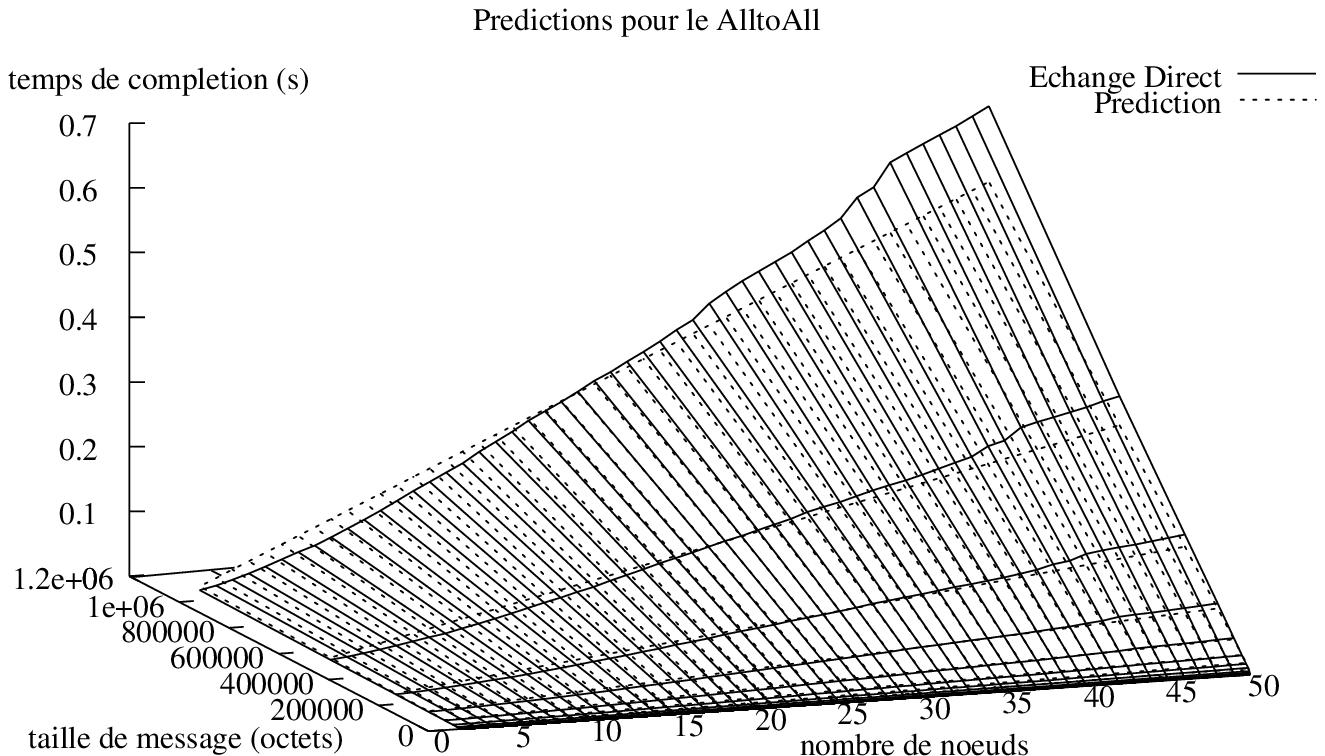}}\tabularnewline
\end{tabular}

\caption{\label{Figure:gamma}Prédictions de performance pour l'algorithme
\emph{All-to-All}}
\end{figure}

Des relations de congestion qui permettent une bonne approximation
des résultats réels présentés en Figure \ref{Figure:Comparison-bounds}
sont $\gamma=\frac{1}{5}$ pour le réseau Fast Ethernet et $\gamma=\frac{3}{2}$
pour le réseau Myrinet. L'application de ce facteur $\gamma$ sur
d'autres expériences où on varie le nombre de processus s'est montrée
assez fiable, surtout pour des grands messages. La prédiction de performance
pour les petits messages reste encore sujette à des facteurs difficiles
à contrôler, comme par exemple les retards dûs à la synchronisation
des processus, la performance du réseau ou les problèmes liés à l'implantation
du protocole TCP. Les prédictions obtenues avec notre modèle de performance
sont présentées en Figure \ref{Figure:gamma}.

\section{\label{sec:Conclusions}Conclusions et Travaux Futurs}

Des travaux récents visent l'optimisation des opérations de communication
collective dans les environnements de type grille de calcul. La solution
la plus répandue est la séparation des communications internes et
externes à chaque grappe, comme le font les systèmes ECO \cite{key-5},
MagPIe \cite{key-6}\cite{key-9}, et LAM-MPI 7 \cite{key-32}, mais
cela n'exclut pas le découpage des communications en plusieurs couches,
pratique efficace démontrée par Karonis \emph{et al.} \cite{key-53}.
Dans les deux cas, la prédiction des performances est un facteur essentiel,
soit pour le réglage fin des paramètres de communication, soit pour
le calcul de la distribution et de la hiérarchie des communications.
Pour cela, il est très important d'avoir des modèles précis des communications
collectives, lesquels seront utilisés pour prédire ces performances. 

Cet article présente notre expérience dans le domaine de la modélisation
des opérations de communication collective. L'efficacité de ces modèles
est analysée à travers la comparaison entre les prédictions de performance
et les résultats réels obtenus pour trois importants patrons de communication
collective : \ggf un vers plusieurs\gdf , \ggf un vers plusieurs
personnalisé\gdf ~et \ggf plusieurs vers plusieurs\gdf . Pour
cela, les expériences ont utilisé deux architectures réseaux différentes,
Fast Ethernet et Myrinet. Nous démontrons que les modèles de communication
sont suffisamment précis pour prédire les performances de ces opérations
collectives sur les deux environnements réseaux, et aussi pour permettre
la sélection des techniques le plus adaptées à chaque situation.

Une contribution importante de cet article est l'effort de modéliser
les opérations de type \ggf plusieurs vers plusieurs\gdf . En général,
ces opérations sont sujettes à des retards importants dûs aux effets
de la congestion du réseau. Dans notre approche, un facteur de congestion
linéaire $\gamma$, obtenu à partir des modèles de performance théoriques,
est utilisé pour prédire les performances de ce type d'opération collective
avec une bonne précision et surtout un coût très bas. Même si notre
modèle de communication \ggf plusieurs vers plusieurs\gdf ~ ne
couvre pas tous les effets de congestion qui peuvent influencer les
résultats réels, en particulier dans le cas des petits messages, il
fournit des indices qui contribuent à la recherche des modèles plus
précis.

Cet article s'encadre dans le contexte de notre recherche sur des
communications collectives adaptées aux environnements de grille.
Nous sommes particulièrement intéressés à la construction automatique
des communications collectives à multiples niveaux, dont la modélisation
des performances, la découverte du réseau et la construction des hiérarchies
de communication sont des aspects essentiels.

\end{document}